\pdfoutput=0
\documentclass[11pt]{article}
\usepackage{axodraw}
\usepackage{epsfig}
\usepackage{amsfonts}
\usepackage{amsmath}
\usepackage{bm,bbm}
\usepackage{cite}
\allowdisplaybreaks[1]

\input pix.sty

\newcommand{\la}[1]{\label{#1}}
\newcommand{\be}{\begin{equation}}
\newcommand{\ee}{\end{equation}}
\newcommand{\ba}{\begin{eqnarray}}
\newcommand{\ea}{\end{eqnarray}}
\newcommand{\rmi}[1]{{\mbox{\scriptsize #1}}}
\newcommand{\fig}{Fig.~}
\newcommand{\figs}{Figs.~}
\newcommand{\eq}{Eq.~}
\newcommand{\eqs}{Eqs.~}
\newcommand{\se}{Sec.~}
\newcommand{\ses}{Secs.~}
\newcommand{\nr}[1]{(\ref{#1})}

\newcommand{\fr}[2]{{\frac{#1}{#2}\,}}

\newcommand{\LambdaIR}{\Lambda_{\mbox{\tiny\rm{IR}}}}
\newcommand{\Lambdaref}{\Lambda_{\mbox{\tiny\rm{0}}}}
\newcommand{\LambdaUV}{\Lambda_{\mbox{\tiny\rm{UV}}}}
\renewcommand{\vec}[1]{{\bf #1}}

\renewcommand{\eq}{eq.~}
\renewcommand{\eqs}{eqs.~}
\renewcommand{\se}{sec.~}
\renewcommand{\ses}{secs.~}
\renewcommand{\fig}{fig.~}
\renewcommand{\figs}{figs.~}

\newcommand{\Nc}{N_{\rm c}}

\newcommand{\Tc}{T_{\rm c}}
\newcommand{\Te}{T_{\rm e}}
\newcommand{\Tnow}{T_{\mbox{\tiny\rm{0}}}}
\newcommand{\fnow}{f_{\mbox{\tiny\rm{0}}}}
\newcommand{\snow}{s_{\mbox{\tiny\rm{0}}}}
\newcommand{\Tstat}{T_{\rm stat}}
\newcommand{\Tmax}{T_{\rm max}}

\newcommand{\rmO}{{\mathcal{O}}}

\newcommand{\dA}{d_\rmii{A}^{ }}
\def\lsi{\raise0.3ex\hbox{$<$\kern-0.75em\raise-1.1ex\hbox{$\sim$}}}
\def\gsi{\raise0.3ex\hbox{$>$\kern-0.75em\raise-1.1ex\hbox{$\sim$}}}
\newcommand{\lsim}{\mathop{\lsi}}
\newcommand{\gsim}{\mathop{\gsi}}

\newcommand{\nB}{n_\rmii{B}}
\newcommand{\rmii}[1]{{\mbox{\tiny\rm{#1}}}}
\newcommand{\rmiii}[1]{{\mbox{\tiny{$\scriptstyle{\rm#1}$}}}}

\newcommand{\Tint}[1]{{\hbox{$\sum$}\!\!\!\!\!\!\!\int\,}_{\!\!\!\!\raise-0.9ex\hbox{$\scriptstyle{#1}$}}}
\newcommand{\Tinti}[1]{{{\Sigma}\!\!\!\!\raise0.3ex\hbox{$\int$}_\rmii{${#1}$}}}
\newcommand{\bi}{\begin{itemize}}
\newcommand{\ei}{\end{itemize}}
\newcommand{\hide}[1]{ }

\newcommand{\deltabar}{\raise-0.02em\hbox{$\bar{}$}\hspace*{-0.8mm}{\delta}}
\newcommand{\ddeltabar}{\raise-0.18em\hbox{$\bar{}$}\hspace*{-0.8mm}{\delta}}

\newcommand{\T}{\rmii{$T$}}

\newcommand{\mpl}{m_\rmii{pl}} 
\def\TAsc(#1,#2)(#3,#4,#5)%
{\SetWidth{2.0}\CArc(#1,#2)(#3,#4,#5)\SetWidth{1.0}}
\def\Lwidth{3}

\def\TAgl(#1,#2)(#3,#4,#5){\SetWidth{2.0}\PhotonArc(#1,#2)(#3,#4,#5){\Lwidth}%
{6.283 #3 mul 360 div #4 #5 sub #4 #5 sub mul sqrt mul Tdensity mul}%
\SetWidth{1.0}}
\def\TLgl(#1,#2)(#3,#4){\SetWidth{2.0}\Photon(#1,#2)(#3,#4){\Lwidth}
{#1 #3 sub #1 #3 sub mul #2 #4 sub #2 #4 sub mul add sqrt Tdensity mul}%
\SetWidth{1.0}}

\renewcommand{\picc}[1]{\;\parbox[c]{60pt}{\begin{picture}(60,30)(0,-10)
\SetWidth{1.0}\SetScale{1.3} #1 \end{picture}}\; }
\def\Lwidth{1.3}

\newcommand{\picu}[1]{\;\parbox[c]{70pt}{\begin{picture}(70,30)(-10,-5)
\SetWidth{1.0}\SetScale{1.3} #1 \end{picture}}\; }

\def\AmplGauge{\picu{%
 \Lgl(10,0)(30,5)%
 \Lgl(10,30)(30,25)%
 \Lgl(30,5)(30,25)%
 \Lgl(30,5)(50,0)%
 \Lgl(30,25)(50,30)%
}}
\def\AmplInfla{\picu{%
 \Lsc(10,0)(30,5)%
 \Lgl(30,4)(30,24)%
 \Lgl(30,5)(50,0)%
 \COval(30,5)(2,2)(0){Black}{Black}%
 \Photon(49.0,33)(29.0,28){1.5}{4}%
 \Photon(49.5,30.5)(29.5,25.5){1.5}{4}%
 \Photon(50.0,28)(30.0,23){1.5}{4}%
 \Photon(11.0,33)(31.0,28){-1.5}{4}%
 \Photon(10.5,30.5)(30.5,25.5){-1.5}{4}%
 \Photon(10.0,28)(30.0,23){-1.5}{4}%
 \CBoxc(30,25.5)(4,9){Black}{White}%
}}
\makeatletter \@addtoreset{equation}{section} \makeatother

\makeatletter
\renewcommand\section{\@startsection {section}{1}{\z@}%
                                   {-5.5ex \@plus -1ex \@minus -.2ex}
                                   {2.3ex \@plus.2ex}%
                                   {\normalfont\large\bfseries}}
\renewcommand\subsection{\@startsection{subsection}{2}{\z@}%
                                     {-3.25ex\@plus -1ex \@minus -.2ex}%
                                     {1.5ex \@plus .2ex}%
                                     {\normalfont\normalsize\bfseries}}
\renewcommand\thesection {\@arabic\c@section}
\renewcommand\thesubsection   {\thesection.\@arabic\c@subsection}
\renewcommand{\@seccntformat}[1]{%
\csname the#1\endcsname.\hspace{1.0em}}
\makeatother

\begin{document}

\flushbottom

\begin{titlepage}

\begin{flushright}
May 2023
\end{flushright}
\begin{centering}
\vfill

{\Large{\bf
     Maximal temperature of strongly-coupled dark sectors
}} 

\vspace{0.8cm}

H.~Kolesova, 
M.~Laine, 
S.~Procacci

\vspace{0.6cm}

{\em
AEC, 
Institute for Theoretical Physics, 
University of Bern, \\ 
Sidlerstrasse 5, CH-3012 Bern, Switzerland \\}

\vspace*{0.6cm}

{\em 
Emails: kolesova@itp.unibe.ch, laine@itp.unibe.ch, procacci@itp.unibe.ch}

\vspace*{0.8cm}

\mbox{\bf Abstract}
 
\end{centering}

\vspace*{0.3cm}
 
\noindent
Taking axion inflation as an example, we estimate the maximal 
temperature ($T_{\rm max}^{ }$)
that can be reached in the post-inflationary universe, 
as a function of the confinement scale of a non-Abelian dark sector 
($\Lambda_{\mbox{\tiny\rm{IR}}}^{ }$). 
Below a certain threshold 
$\Lambda_{\mbox{\tiny\rm{IR}}}^{ } < \Lambda_{\mbox{\tiny\rm{0}}}^{ }
\sim 2\times 10^{-8}_{ } m_{\rm pl}^{ }$, 
the system heats up to 
$T_{\rm max}^{ } \sim \Lambda_{\mbox{\tiny\rm{0}}}^{ } > T_{\rm c}^{ }$, 
and a first-order thermal phase transition takes place. 
On the other hand, if 
$\Lambda_{\mbox{\tiny\rm{IR}}}^{ } > \Lambda_{\mbox{\tiny\rm{0}}}^{ }$, 
then $T_{\rm max}^{ } \sim \Lambda_{\mbox{\tiny\rm{IR}}}^{ } < T_{\rm c}^{ }$:
very high temperatures can be reached, but there is no phase transition.  
If the inflaton thermalizes during heating-up 
(which we find to be unlikely), 
or if the plasma includes light degrees of freedom, then heat capacity 
and entropy density are larger, and $T_{\rm max}^{ }$ is 
lowered towards $\Lambda_{\mbox{\tiny\rm{0}}}^{ }$. The heating-up dynamics 
generates a gravitational wave background. 
Its contribution to $N^{ }_{\rm eff}$ at GHz frequencies, 
the presence of a monotonic $\sim f_{\mbox{\tiny\rm{0}}}^3$ shape at 
$(10^{-4}_{ } - 10^2_{ })\,$Hz frequencies, and the frequency domain
of peaked features that may originate 
via first-order phase transitions, are discussed. 

\vfill


\end{titlepage}

\tableofcontents

%
\section{Introduction}
\la{se:intro}

As the nature of dark matter remains unresolved and non-standard
ideas have become an accepted part of the speculation, one of the 
avenues is to envisage the existence of a whole dark sector. 
There is a great variety of possibilities for the field content 
of the dark sector and for its interactions with the visible one. 
Yet any dark sector surely couples to gravity, 
and then it is natural to assume that it connects  
to inflationary dynamics as well. 

If the dark sector consists 
of a non-Abelian Yang-Mills theory, 
so that gauge invariant operators have 
dimension 4, then its interactions with the Standard Model 
can be very weak, possibly
even suppressed by the Planck mass squared
(cf.,\ e.g.,\ refs.~\cite{ulb,soni,dark} and references therein). 
Weak interactions between the dark and visible sectors 
allow the dark sector temperature to differ
from the Standard Model one. We would like to know how high
the dark sector temperature can be, 
as this affects several phenomena, such as the spectrum
of gravitational waves that gets generated; 
the efficiency with which dark matter candidates
can be produced;
and the kind of thermal phase
transitions that can be encountered.

When we talk about thermodynamic notions, it is a relevant
question under which conditions 
the temperature can be defined at all. 
For a non-Abelian Yang-Mills theory, the generic equilibration
rate, originating from kinematically unconstrained $2\to 2$
scatterings, is of order 
$\Gamma^{ }_g \sim \alpha^2 T^{ }_\rmi{dark}$, 
where $\alpha \equiv g^2/(4\pi)$
is the gauge coupling. 
An upper bound on the temperature is obtained by comparing 
this with the Hubble rate
of a radiation-dominated expanding universe, 
$H \sim \max\{T^2_\rmi{dark},T^2_\rmi{visible}\} / \mpl^{ }$, 
where 
$\mpl^{ }$ 
is the Planck mass. 
For $T^{ }_\rmi{dark} < \alpha^2 \mpl^{ } $, 
the equilibration rate exceeds the Hubble rate, 
i.e.~$\Gamma^{ }_g > H$.
If we consider dark sectors 
with $\alpha \sim 0.3$, it is therefore in principle meaningful 
to discuss temperatures almost up to the Planck scale. 
On the other hand, if the Hubble rate is dominated by 
a temperature-independent part, like a vacuum energy density,
there is also a lower bound on~$T^{ }_\rmi{dark}$. 
We will return to 
{\it a posteriori} comparisons of the equilibration
and Hubble rates. 

In order to carry out a concrete discussion, 
we adopt a specific inflationary scenario that can indeed be 
argued to thermalize efficiently (cf.\ \se\ref{ss:equil}), 
namely that of non-Abelian axion-like (or natural) inflation~\cite{ai}.
The parameters of the inflaton potential are fixed from 
standard ``cold inflation'' predictions, 
to match Planck data~\cite{planck}. The heating-up dynamics is characterized
by the gauge coupling $\alpha$ that does not affect inflationary
predictions at leading order. 
Apart from the self-interactions of the Yang-Mills
plasma, $\alpha$ also parametrizes the interactions
between the dark sector and the inflaton, and for this we adopt
the form of the pseudoscalar operator, 
\be
 \mathcal{L} 
 \; \supset \;
 \frac{ 1 }{2} \partial^\mu\varphi\, \partial_\mu\varphi
 - V^{ }_0(\varphi)  
 - \frac{\varphi \,  \chi}{f^{ }_a}
 \;, 
 \quad
 \chi \;\equiv\;
 \frac{ 
 \alpha \,
 \epsilon^{\mu\nu\rho\sigma}_{ }
 F^{c}_{\mu\nu} F^{c}_{\rho\sigma}
 }{16\pi}
 \;, \la{L}
\ee
where $\varphi$ is the inflaton field, 
$
 F^c_{\mu\nu} 
$ 
is the Yang-Mills field strength, 
$ c $ 
is a colour index, 
and $f^{ }_a$ is the axion decay constant. 
The advantage of this interaction term is that concrete
(even if so far incomplete)
information is available about the friction and 
mass corrections that it leads to. We normally
reparametrize $\alpha$ through a dark confinement scale
$\LambdaIR^{ }$, cf.\ \eq\nr{alpha}. Furthermore, 
for simplicity, we denote $ T \equiv T^{ }_\rmi{dark}$ in the following, 
and assume that the effect of the visible sector can be neglected
in the period of time that we are interested in.\footnote{%
  If the dark sector consists of a relativistic plasma, 
  this assumption is roughly equivalent to 
  $T^{ }_\rmii{dark} > T^{ }_\rmii{visible}$.
  There are concrete scenaria where it has been argued that
  the dark sector indeed heats up first and injects subsequently
  a part of its energy density into the visible one, so that
  the dark sector could be hotter than the Standard Model, 
  see e.g.\ refs.~\cite{bk,seq_reheat,fk} and references therein.
   }

Our presentation is organized as follows. 
After an exposition of our general 
setup (cf.~\se\ref{se:setup}), we first introduce
the concept of a stationary temperature. The latter permits
for a simple qualitative estimate of the energy density 
that the Yang-Mills plasma obtains during inflation
(cf.~\se\ref{ss:Tstat}).
For a quantitative understanding, we then proceed to
a numerical solution of the maximal temperature, which
is somewhat higher than the stationary one (cf.~\se\ref{ss:Tmax}). 
After elaborating upon physical implications
for gravitational waves
(cf.~\se\ref{se:phys}), we turn to a summary
and outlook (cf.~\se\ref{se:concl}).

%
\section{Inflationary setup}
\la{se:setup}

%
\subsection{Evolution equations}
\la{ss:eqs}

Given the fast thermalization rate 
of non-Abelian gauge theory (the arguments for this are revisited 
in~\se\ref{ss:equil}), 
we carry out our discussion assuming that
the notion of a local temperature-like quantity can be defined. 
The degrees of 
freedom are then the average inflaton field, $\bar\varphi$, and the plasma 
temperature, $T$. 
The equations governing their evolution 
can be written as
(a justification from energy conservation follows 
below \eq\nr{eom_field_2}) 
\ba
 \ddot{\bar\varphi} +
 (3 H + \Upsilon)\dot{\bar\varphi}
 +  V^{ }_\varphi 
 & \simeq & 
  0
 \;, \la{eom_field} \\[2mm] 
 \dot{e}^{ }_r
 + 3 H \bigl( e^{ }_r + p^{ }_r - T  V^{ }_\T \bigr)
 - T \dot{V}^{ }_\T  
 & \simeq &
 \Upsilon^{ }
 \dot{\bar\varphi}^2
 \;, \la{eom_plasma}
\ea
where $e^{ }_r$ and $p^{ }_r$ denote
the energy density and pressure of radiation. 
Furthermore $ H \equiv \dot{a}/a$ is the Hubble rate;
$V$ is the inflaton potential;\footnote{%
 How it differs from the tree-level potential $V^{ }_0$
 in \eq\nr{L} is discussed in \se\ref{ss:mass}.}
$V^{ }_x \equiv \partial^{ }_x V$;
and $\Upsilon$ is a friction coefficient, 
which transfers energy from the inflaton 
to radiation degrees of freedom (cf.\ \se\ref{ss:upsilon}). 
If we set $T\to 0$ and $\Upsilon\to 0$ as initial conditions, 
the plasma remains at zero
temperature, and we return back to normal cold inflation. 

Denoting the pressure and energy density carried by the inflaton by
\be
 p^{ }_\varphi
 \; \equiv \; \frac{\dot{\bar\varphi}^2}{2} - V 
 \;, \quad
 e^{ }_\varphi
 \; \equiv \; \frac{\dot{\bar\varphi}^2}{2} + V - T V^{ }_\T 
 \;, \la{e_phi}
\ee
and multiplying \eq\nr{eom_field} by $\dot{\bar\varphi}$, 
the evolution equation for $\bar\varphi$
can equivalently be expressed as 
\be
 \dot{e}^{ }_\varphi
 + 3 H \dot{\bar\varphi}^2 + T \dot{V}^{ }_\T 
 \; \simeq \; 
 - \Upsilon \dot{\bar\varphi}^2
 \;. \la{eom_field_2} 
\ee
Summing together \eqs\nr{eom_plasma} and \nr{eom_field_2} yields
the overall energy conservation equation, \linebreak 
$ 
 \dot{e} + 3 H(e+p) = 0
$, 
where 
$
 e \equiv e^{ }_\varphi + e^{ }_r
$
and 
$
 p \equiv p^{ }_\varphi + p^{ }_r
$.
In the same notation
the Friedmann equation reads 
$
 H^2_{ } \; = \; 8\pi e / (3\mpl^2 )  
$, 
where $\mpl^{ } = 1.22091 \times 10^{19}_{ }$~GeV.

\vspace*{3mm}

As we will see in \se\ref{ss:Tmax}, the system just defined 
can cross a first-order phase transition.
In this case, 
\eq\nr{eom_plasma} needs to be supplemented by 
another equation, valid when the system is in 
a mixed phase. The technical reason is that 
in a mixed phase, 
the temperature stays constant at $T = \Tc^{ }$, 
so that $\dot{T} = 0$. 
At the same time, the energy density
has a discontinuity, $e^{ }_r (\Tc^+) - e^{ }_r(\Tc^-) > 0$, 
so that 
$ 
 \partial^{ }_\T e^{ }_r |^{ }_{T = \Tc}
$
diverges. 
Therefore a naive evaluation of the time derivative
is ambiguous,  
$\dot{e}^{ }_r = \dot{T} \partial^{ }_\T e^{ }_r 
 = \mbox{``} 0\times\infty \mbox{''} $. 

In the real world, a mixed phase can incorporate complicated 
physics (bubble nucleations, sound wave dynamics, turbulence).
However, the
overall picture should be well captured by an adiabatic
approximation. In this treatment, we re-parametrize 
$e^{ }_r(t) |^{ }_{T = \Tc}$ through a volume fraction, $u$, as
\ba
 e^{ }_r(t) & \equiv &
              e^{ }_r(\Tc^+) \, u(t) 
            + e^{ }_r(\Tc^-) \, [1 - u(t)]
 \;, \quad
 0 \le u \le 1
 \;  \la{u_def}
 \\ 
 \Rightarrow \; 
 \dot{e}^{ }_r(t) & = & 
 \dot{u}(t) \, 
 \bigl[ \,
  e^{ }_r(\Tc^+) - e^{ }_r(\Tc^-)
 \, \bigr]
 \;. 
\ea
In contrast, the pressure $p^{ }_r$ is continuous at $T = \Tc^{ }$,
since it equals minus the free energy density, 
and therefore independent of $u$.
Thereby \eq\nr{eom_plasma} gets replaced with 
\be
 \dot{u} \, 
 \bigl[ \,
  e^{ }_r(\Tc^+) - e^{ }_r(\Tc^-)
 \, \bigr]
 + 3 H \bigl( e^{ }_r + p^{ }_r - T  V^{ }_\T \bigr)
 - T \dot{V}^{ }_\T  
 \; \simeq \;
 \Upsilon^{ }
 \dot{\bar\varphi}^2
 \;. \la{eom_plasma_2}
\ee
We note that the potential $V$ and its derivatives are well-defined, 
since the inflaton field {\em does not} undergo 
any phase transition in our setup.

\begin{figure}[t]

\hspace*{-0.1cm}
\centerline{%
   \epsfysize=5.0cm\epsfbox{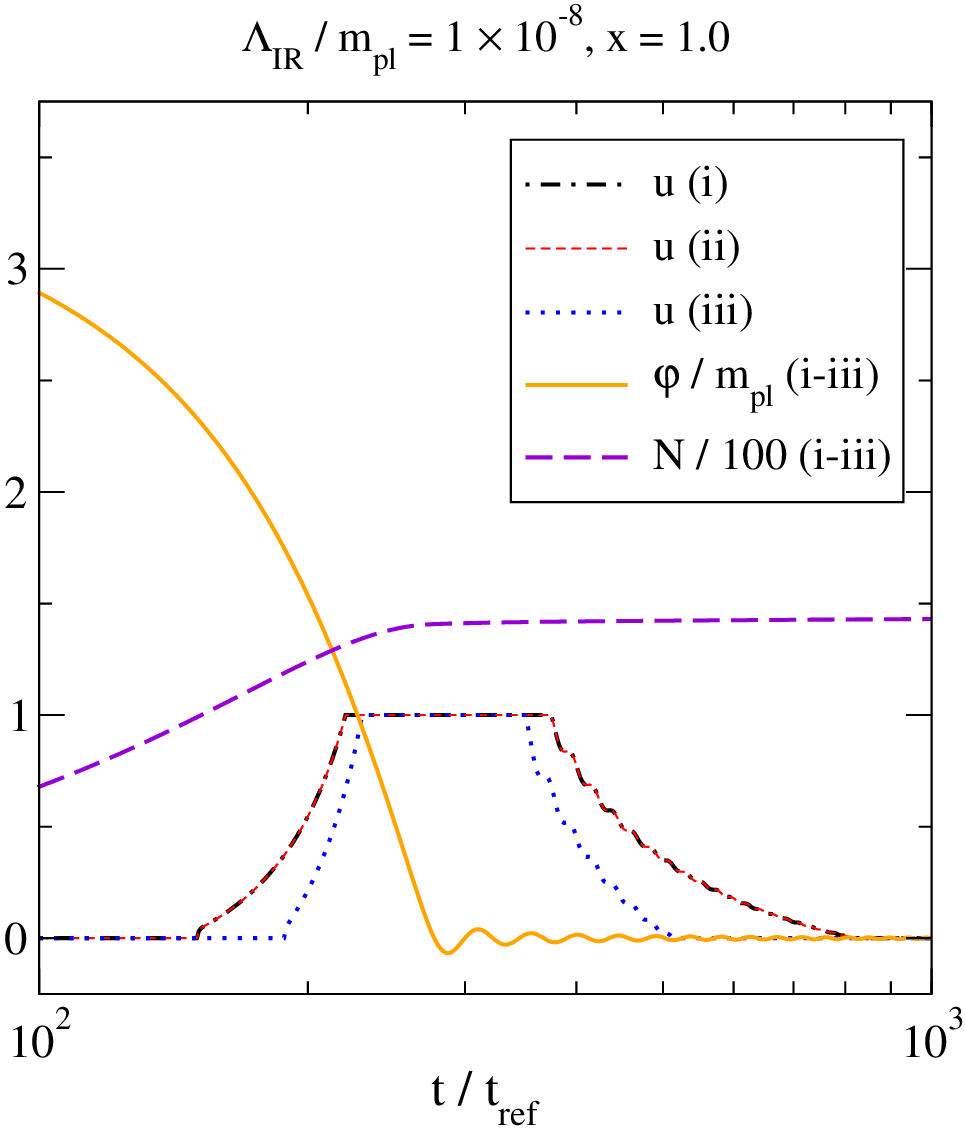}
   ~~~\epsfysize=5.0cm\epsfbox{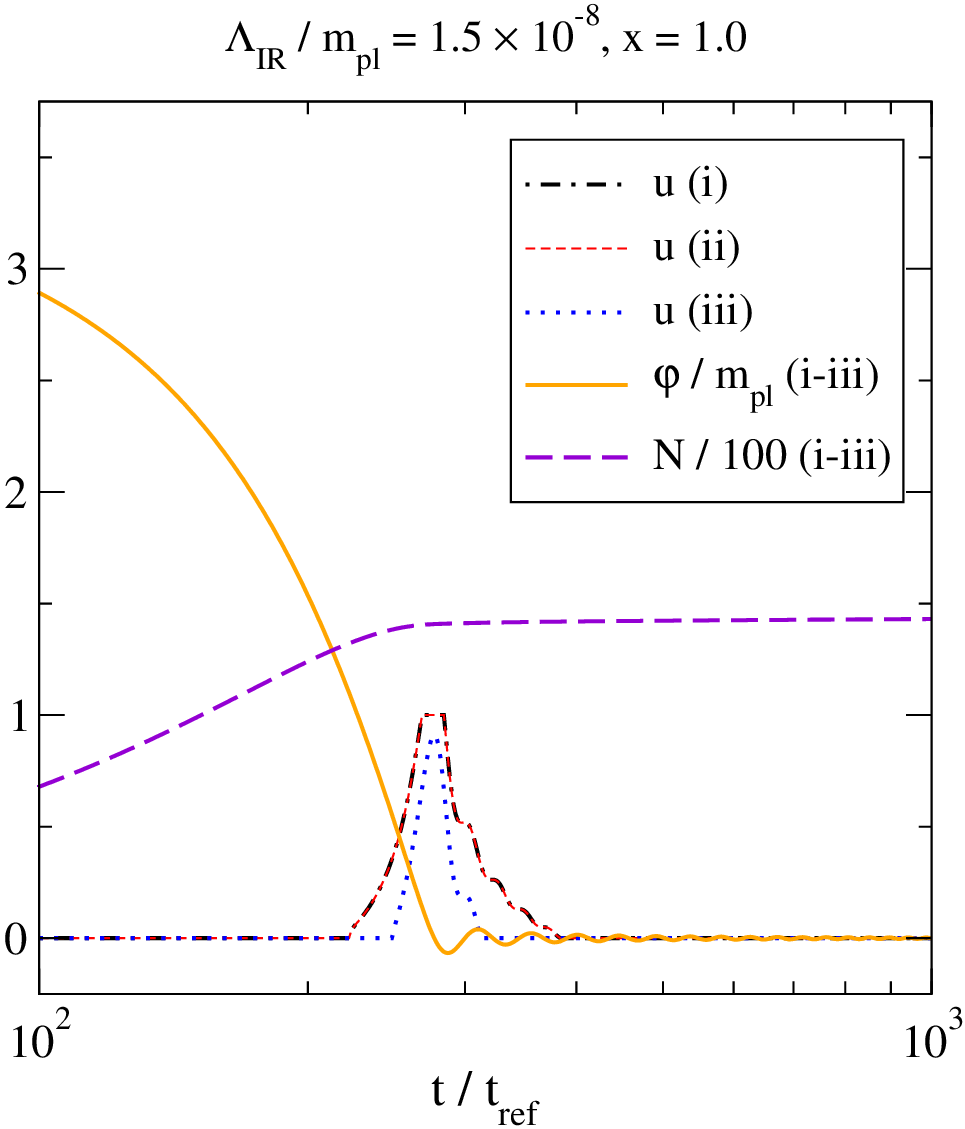}
   ~~~\epsfysize=5.0cm\epsfbox{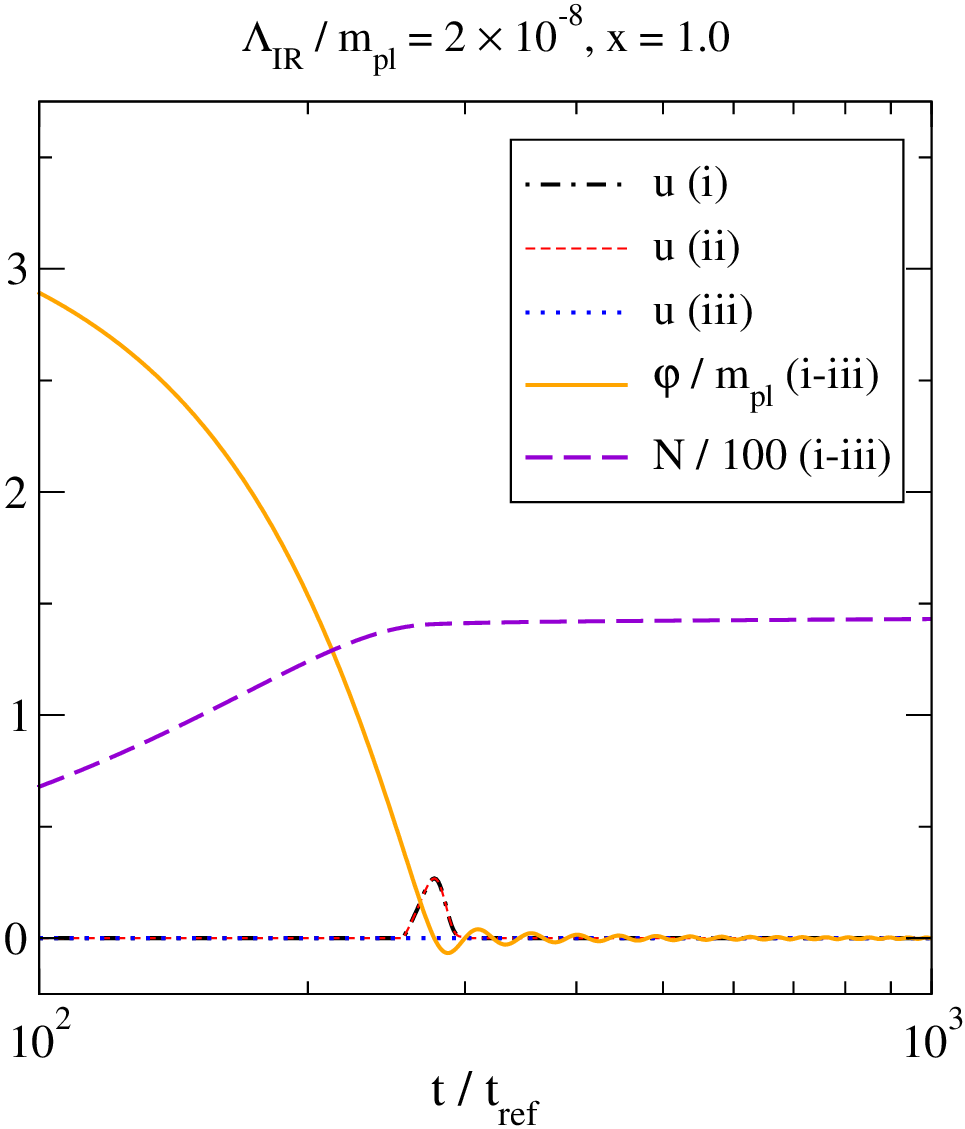}
}

\caption[a]{\small
  Examples in which the critical point is reached
  and surpassed (left, middle) or only reached (right). 
  Here $u$ denotes the volume fraction of the deconfined phase 
  (cf.\ \eq\nr{u_def}), 
  $N$ the number of $e$-folds from $t = t^{ }_\rmi{ref}$ (cf.\ \eq\nr{tref}), 
  and $x$ a parameter introduced in \eq\nr{alpha}.
  The cases (i--iii) are defined in \se\ref{ss:Tstat}.  
  The temperature evolution of the left and right
  solution is shown in \fig\ref{fig:soln}.
}

\la{fig:mixed}
\end{figure}

The solution of the differential equations needs now to be 
complemented by a monitoring of $T$ and $u$
(cf.\ \fig\ref{fig:mixed} for an illustration). 
If we are solving \eq\nr{eom_plasma}, and notice that 
$ 
 T \to \Tc^-
$
(respectively $T \to \Tc^+$), 
then we need to go over into \eq\nr{eom_plasma_2}, 
with the initial condition
$
 u = 0
$
(respectively $u = 1$).
If we are solving \eq\nr{eom_plasma_2}, and notice that
$
 u \to 0
$
($u \to 1$), 
then we need to go over into \eq\nr{eom_plasma}, 
with the initial condition 
$
 e^{ }_r = e^{ }_r(\Tc^-)
$
($e^{ }_r = e^{ }_r(\Tc^+)$).
It is possible that the system enters and exits the mixed
phase from the same side (for instance, if $\Tmax^{ } = \Tc$), 
or from different sides (if the transition is passed through
on the way towards higher or lower temperatures).

%
\subsection{Non-perturbative thermodynamic functions for a radiation plasma}
\la{ss:s_r}

As an essential 
ingredient to \eqs\nr{eom_plasma} and \nr{eom_plasma_2}, we need
the thermodynamic energy density  and pressure of the 
radiation plasma, $e^{ }_r$ and $p^{ }_r$. 
These are often parametrized through degrees of freedom
$g^{ }_*$ or $h^{ }_*$, as 
$e^{ }_r \equiv g^{ }_* \pi^2 T^4/30$ and 
$e^{ }_r + p^{ }_r  = T s^{ }_r \equiv 
h^{ }_* 2 \pi^2 T^4/45$, 
where $s^{ }_r$ is the entropy density. 
If the plasma is very weakly coupled, 
$g^{ }_*$ is to a good approximation constant and $h^{ }_* \simeq g^{ }_*$, 
however our focus is on self-interacting plasmas, where the
interactions can become strong as well.\footnote{%
 In this section and in \se\ref{ss:mass}, we make use of 
 non-perturbative information, so that the coupling can be 
 arbitrarily strong, whereas for the friction coefficient
 discussed in \se\ref{ss:upsilon}, 
 reliable non-perturbative information is not available. 
 Then we extrapolate weak-coupling predictions, strictly 
 speaking only applicable for $\alpha < 0.3$, 
 to a strongly coupled regime, introducing
 an estimate of the corresponding error
 along the way.
 } 
In the latter case, 
$g^{ }_*$ and $h^{ }_*$ decrease rapidly at low temperatures, 
and their complete functional forms are needed. 

It turns out to be convenient to parametrize the thermodynamic
information through the entropy density, $s^{ }_r$. On one hand, this is 
because we need $s^{ }_r$ for \eq\nr{master}; on the other, 
because $s^{ }_r$ can be precisely  studied
with lattice simulations~(cf.\ refs.~\cite{eos_jp,eos} and
references therein). 
Denoting by $\Tc^{ }$ the critical temperature, 
and setting $\Nc^{ } = 3$ from now on, 
the results of the deconfined phase of a Yang-Mills plasma 
can be represented as~\cite{eos}\footnote{%
 Ref.~\cite{eos} gives $T = 3.433\Tc^{ }$ as the transition
 point between the two functional forms; we have replaced this
 with the value at which the curves cross each other, 
 with the approximate coefficients at our disposal. 
 }
\be
 \frac{s^{ }_r}{T^3} \bigg|^{ }_\rmi{lattice}
 \;\approx\;
 \left\{ 
 \begin{array}{ll}
 \displaystyle
 6.9829 - \frac{1.0348}{\ln(T/\Tc)} 
 & 
 \;, \quad 
 T \ge 3.222 \Tc^{ } \\[3mm] 
 \displaystyle
 \frac{1.7015 + 77.757 \ln (T/\Tc) + 232.33 \ln^2(T/\Tc)}
      {1.0 + 19.033 \ln(T/\Tc) + 32.200 \ln^2(T/\Tc)}
 & 
 \;, \quad 
 \Tc^{ } < T < 3.222 \Tc^{ } 
 \end{array}
 \right. 
 \;. \la{s_highT}
\ee
For $T/\Tc \to \infty$ this agrees within 0.5\% with the 
Stefan-Boltzmann value 
\be
 \frac{s^{ }_r}{T^3} \biggr|^{ }_\rmi{free} = 
 \frac{ 2\pi^2 \times 16 }{ 45 } = 7.018
 \;. \la{sb}
\ee

The determination of $s^{ }_r/T^3$ is more difficult in the 
confined phase, as the results soon become exponentially small. 
Fitting to the tabulated results from ref.~\cite{eos},\footnote{%
 The first number can be found in the text, not the table. 
 The last number appears to contain a typo in the table of ref.~\cite{eos};
 we have reconstructed the correct value from the 
 $e^{ }_r$ and $p^{ }_r$ given on the same line.
 } {\it viz.}\
\be
  \begin{array}{ll}
       s^{ }_r / T^3   & T / \Tc^{ }  \\ \hline
       0.37(15)  & 1.0^-_{ }    \\
       0.31(11)  & 0.980        \\ 
       0.108(23) & 0.904        \\ 
       0.001(4)  & 0.660        
  \end{array}
 \;, 
\ee
which appear to be consistent with ref.~\cite{confined}, we model the
low-$T$ region with the ansatz
\be
 \frac{s^{ }_r}{T^3} \bigg|^{ }_\rmi{lattice} 
 \; \stackrel{ T < \Tc^{ } }{\simeq} \; 
 a \, \biggl( \frac{T}{\Tc^{ }} \biggr)^{b}_{ }
 \, \exp \biggl( - \frac{c\, \Tc^{ }}{T} \biggr)
 \;, \quad a = 45.8
 \;, \quad b = 6.81
 \;, \quad c = 4.80
 \;. \la{s_lowT}
\ee
The transition is of the first order, 
so that $s^{ }_r/T^3$ displays a discontinuity at $T = \Tc^{ }$. 
For the conversion between $\Tc^{ }$ 
and $\LambdaIR^{ }$, 
we estimate 
$\Tc^{ }\simeq 1.24\LambdaIR^{ }$~\cite{Tc}. 

Given $s^{ }_r = {\rm d}p^{ }_r / {\rm d}T$, the other thermodynamic 
functions can be obtained as 
\be
 p^{ }_r(T) - p^{ }_r (0) = 
 \int_0^T \! {\rm d}T'\, T'^3_{ } \, \biggl( \frac{s^{ }_r}{T'^3} \biggr)
 \;, \quad
 e^{ }_r(T) - e^{ }_r (0) = 
 T s^{ }_r - \bigl[ p^{ }_r(T) - p^{ }_r (0) \bigr]
 \;. 
\ee
Furthermore, 
in order to evaluate $\dot{e}^{ }_r = \dot{T} c^{ }_r$, 
we need the heat capacity 
$c^{ }_r = \partial^{ }_\T e^{ }_r = T \partial^{ }_\T s^{ }_r $.
To keep $c^{ }_r$ continuous, 
we have moved the matching point in \eq\nr{s_highT}
to $T = 4.863 \Tc^{ }$ when evaluating $ \partial^{ }_\T s^{ }_r $.
At low temperatures, in turn, \eq\nr{s_lowT} implies
\ba
 \frac{p^{ }_r(T) - p^{ }_r(0) }{T^4}
 & \stackrel{ T < \Tc^{ } }{\simeq} & 
 a\, c^{b}_{ }
 \biggl( \frac{c \Tc^{ }}{T} \biggr)^4_{ }
 \Gamma\biggl( -b-4,\frac{c\,\Tc^{ }}{T} \biggr)
 \;, \la{p_lowT} \\ 
 \frac{c^{ }_r}{T^3} 
 & \stackrel{ T < \Tc^{ } }{\simeq} &
 a\, \biggl[ 
  \bigl( b + 3 \bigr)
  \biggl( \frac{T}{\Tc^{ }} \biggr)^b_{ }
  + c 
  \biggl( \frac{T}{\Tc^{ }} \biggr)^{b-1}_{ }
 \biggr]
 \, \exp \biggl( - \frac{c\, \Tc^{ }}{T} \biggr)
 \;, \la{c_lowT}
\ea
where 
$
 \Gamma(s,x) = \int_x^\infty \! {\rm d}t \, t^{s-1}_{ } \, e^{-t}_{ }
$
is an incomplete gamma function. 

Restricting to the SU(3) plasma is a special case, 
however this is the system for which the most reliable non-perturbative 
information is available. In addition, it entails
a weak first-order transition, 
which is typical of many other thermal systems. 

%
\subsection{Perturbative thermodynamic functions for a thermalized inflaton}
\la{ss:s_varphi}

As the system heats up, 
the inflaton field might equilibrate as well 
(see, however, the discussions in \ses\ref{ss:equil} and \ref{ss:Tmax}). 
Around the minimum of the potential, 
the inflation is a weakly coupled massive scalar field, 
whose interactions are suppressed by powers of $1/f^{ }_a$. 
Then the effective
potential~$V$ contains a temperature dependent part, which 
contributes to the thermodynamic functions 
$p^{ }_\varphi$ and $e^{ }_\varphi$ according to 
\eq\nr{e_phi}.

The starting point for the evaluation
of $p^{ }_\varphi$ and $e^{ }_\varphi$ is the 1-loop expression
for a thermal effective potential, 
\be
 V_\rmi{eff}^{(1)}
 \; = \; 
 \int_\vec{p} \biggl[ 
    \frac{\epsilon^{ }}{2} 
+ T \ln \bigl( 
        1 - e^{-\epsilon/T}_{ }
    \bigr)
 \biggr]^{ }_{\epsilon = \sqrt{p^2 + m^2}}
 \;.
\ee
We omit the $T$-independent vacuum part
of $ V_\rmi{eff}^{(1)} $ in the following. 
Changing variables to a form convenient for a numerical evaluation, 
the pressure and energy density 
from \eq\nr{e_phi} then obtain the contributions
\ba
 - p^{ }_\varphi & \supset & V \;\supset\;
 V^{ }_0 + 
 \frac{T}{2\pi^2}
 \int_m^\infty \! {\rm d}\epsilon\, \epsilon \sqrt{\epsilon^2 - m^2}\,
 \ln\bigl( 1 - e^{-\epsilon / T}_{} \bigr)
 \;, \la{p_varphi} \\
 e^{ }_\varphi & \supset & V - T V^{ }_\T \;\supset\;
 V^{ }_0 
 + \frac{1}{2\pi^2}
 \int_m^\infty \! {\rm d}\epsilon\, \epsilon^2_{ } \sqrt{\epsilon^2 - m^2}\,
 \nB^{ }(\epsilon)
 \;,
\ea
where 
$
 \nB^{ }(\epsilon) \equiv 1 / (e^{\epsilon/T} - 1)
$ is the Bose distribution. The tree-level potential $V^{ }_0$ should contain
no $T$-dependence, so that we have set $V^{ }_{0,\T}$ to vanish, 
however this requires some discussion of thermal mass 
corrections, to which we return in \se\ref{ss:mass}. 
Finally, \eqs\nr{eom_plasma} and \nr{eom_field_2} contain 
the contribution of $\varphi$ to the heat capacity, 
\be
- T \dot{V}^{ }_\T \;\supset\;
 \frac{\dot{T}}{2\pi^2 T^2}
 \int_m^\infty \! {\rm d}\epsilon\, \epsilon^3_{ } \sqrt{\epsilon^2 - m^2}\,
 \nB^{ }(\epsilon)\, \bigl[ 1 + \nB(\epsilon) \bigr]
 \;. \la{c_varphi}
\ee
In the massless limit, i.e.\ $m \ll \pi T$, 
\eqs\nr{p_varphi}--\nr{c_varphi} amount
to the substitutions $g^{ }_* \to g^{ }_* + 1$
and $h^{ }_* \to h^{ }_* + 1$ in the number
of effective degrees of freedom.\footnote{%
 A numerical evaluation at low or intermediate temperatures may be 
 facilitated by representations in terms of modified Bessel functions, 
 $
  p^{ }_\varphi \supset \frac{m^2T^2}{2\pi^2}\sum_{n=1}^{\infty} 
  \frac{1}{n^2}K^{ }_2(\frac{n m}{T})
 $, 
 $ 
  e^{ }_\varphi \supset \frac{m^2T^2}{2\pi^2}\sum_{n=1}^{\infty}
 \bigl\{ 
   \frac{1}{n^2} K^{ }_2(\frac{n  m}{T})
  + \frac{m}{2 n T}
    \bigl[ K^{ }_1(\frac{n  m}{T}) + K^{ }_3(\frac{n  m}{T})\bigr]
 \bigr\}
 $, 
 $ 
  -T\dot{V}^{ }_\T \supset
 \frac{m^4_{ } \dot{T}}{4\pi^2 T}\sum_{n=1}^{\infty}
 \bigl\{ 
  K^{ }_2(\frac{n  m}{T})
  +
  K^{ }_4(\frac{n  m}{T})
 \bigr\}
 $.  
 }  

%
\subsection{Friction coefficient}
\la{ss:upsilon}

A key role for the heating-up dynamics, 
according to \eq\nr{eom_plasma}, 
is played by the friction coefficient~$\Upsilon$. If $\Upsilon = 0$, 
like in standard cold inflation computations, 
there is no source term for the temperature evolution, 
and any possible initial temperature just redshifts away. 

It has been realized, however, that 
the assumption $\Upsilon = 0$ is mathematically 
troublesome. The problem is that even if the $T = 0$
solution represents a fixed point, it can be an unstable one. Just
a small perturbation may drive the system to another fixed point, 
where $T > 0$ and $\Upsilon > 0$~\cite{warm_axion1,fixed_pt}. 
The properties of this thermal fixed point 
constitute the topic of \se\ref{ss:Tstat}. 

In general, $\Upsilon$ is 
a function of the frequency, $\omega$,  
at which the system is probed~\cite{warm}.
Then the full equation
of motion does not have a local form, but rather
contains a dispersive integral over the medium response. 

A local evolution equation is obtained around the global minimum, where the
frequency scale can be replaced by the corresponding
mass scale, $\omega\to m$.
Before the system settles to the global minimum, 
the situation may be intuitively probed by replacing 
the frequency scale by the curvature of the potential, 
$
 \omega\to 
 \omega_\rmi{dyn}^{ } \; \equiv \; \sqrt{\max(0,V^{ }_{\varphi\varphi})}
$~\cite{scan}. 
Then, if $ V^{ }_{\varphi\varphi} \le 0$, 
temperature is the only scale at early stages of inflation. 
But, as mentioned, $T=0$ is an unstable
fixed point in this setup. 
As the focus of the current study is the heating-up
period, we will adopt the replacement $\omega\to m$ throughout, 
with the understanding that at early stages of inflation
this is just a recipe. 

Like the thermodynamic functions in 
\se\ref{ss:s_r}, the determination
of $\Upsilon$ requires lattice simulations. Lattice simulations
come in two different variants. At very high temperatures, 
$T \gg \LambdaIR^{ }$, so-called classical real-time simulations
can be employed, and this is the method used for estimating
$\Upsilon^{ }_\rmiii{IR}$ in \eq\nr{Upsilon_IR}~\cite{mt}, as well as 
the shape in \eq\nr{Upsilon_full}~\cite{clgt}. 
However, when $T \lsim \LambdaIR^{ }$, 
these effective-theory type setups should be replaced
by full four-dimensional lattice simulations. Unfortunately, 
extracting real-time information from the latter is 
exponentially hard (cf.,~e.g.,~refs.~\cite{cuniberti,test}), 
even if exploratory
studies for determining~$\Upsilon^{ }_\rmiii{IR}$ 
have been launched~\cite{explore,sphaleron_4d}. 
For this reason, our estimates contain 
a systematic error, reflected by the $x$-dependence
introduced in \eq\nr{alpha}. 

Now, for a qualitative understanding, 
it is often sufficient to consider limiting ``thermal''  
($\omega \ll T$) or ``vacuum'' ($\omega \gg T$)
frequencies~\cite{warm}.\footnote{%
 %
 %
 In the numerical solutions, we use the full interpolating
 function as estimated in ref.~\cite{clgt}, {\it viz.}  
 \be
 \Upsilon  
 \; 
 \simeq
 \;  
 \frac{\dA \alpha^2}{f_a^2}
 \biggl\{  
 \kappa\, ( \alpha \Nc^{ } T)^3_{ } \, 
 \frac{1 + 
   \frac{\omega^2}
        { (  c^{ }_\rmiii{IR} \alpha^2 \Nc^2 T )^2_{ } }
 }{ 1 + 
   \frac{\omega^2}{ (  c^{ }_\rmiii{M} \alpha \Nc^{ } T )^2_{ } }
 }
 + 
 \Bigl[ 1 + 2 \nB^{ }\Bigl( \frac{\omega}{2} \Bigr)\Bigr]
 \frac{ \pi\omega^3 }{(4\pi)^4} 
 \biggr\}^\rmii{$ c^{ }_\rmiii{IR} \; \simeq \; 106 $ }
         _\rmii{$ c^{ }_\rmiii{M}  \; \simeq \; 5.1 $ } 
 \;. \la{Upsilon_full}
\ee
 %
 %
 } 
Then
\be 
 \Upsilon \; \stackrel{\omega \ll T}{\simeq} \; 
 \Upsilon^{ }_\rmiii{IR}   \; \equiv \;  
 \frac{\dA \alpha^2 \kappa\, ( \alpha \Nc^{ })^3_{ } T^3_{ }}{f_a^2}
 \;, \quad
 \Upsilon \; \stackrel{\omega \gg T}{\simeq} \; 
 \Upsilon^{ }_\rmiii{UV}  \; \equiv \;   
 \frac{\dA \alpha^2}{4 f_a^2}
 \frac{ \omega^3 }{(4\pi)^3} 
 \;, \la{Upsilon_IR}
\ee
where $\dA \equiv \Nc^2 - 1$ and 
$ 
 \kappa  \simeq 1.5
$.
The latter represents the vacuum decay width for the process
$\varphi\to g g$.
For the gauge coupling, 
we adopt a leading-logarithmic running value, 
representative of a Yang-Mills plasma, 
\be
 \alpha \;\simeq\; \frac{6 \pi}{11 \Nc^{ }}
 \ln^{-1}_{ }\biggl[ \frac{\sqrt{(x\,2\pi\LambdaIR^{ })^2
  + (2\pi T)^2 + \omega^2}}
                   {\LambdaIR^{ }}
             \biggr]
 \;. \la{alpha}
\ee
The first term in the square root serves as an (arbitrary)
infrared (IR) regulator, so that 
any value of $T$ or $\omega$ can be inserted;
we will check the IR sensitivity of the results by varying 
the parameter $x$ in the range $x\in (0.2 ... 2.0)$.
Nevertheless the expression is guaranteed to be 
physically meaningful only for 
$\max\{ 2\pi T,\omega \} \gg \LambdaIR^{ }$, 
so that $\alpha \ll 0.3$.

%
\subsection{Mass correction}
\la{ss:mass}

Apart from a friction coefficient, 
the Yang-Mills plasma in general induces a mass correction to the 
inflaton field~\cite{warm}. This is again a function 
of the frequency, $\omega$. 
In principle we could carry out a discussion similar to $\Upsilon$, 
adapting a weak-coupling computation from 
$T \gg \LambdaIR^{ }$~\cite{scan} to a strongly coupled regime
through a modelling of $\alpha$. However, 
partial non-perturbative lattice information is also available,
so we would like to make use of it, even if it is 
not exactly what is needed. Let us explain the issues.

The lattice simulations are usually
viewed in the context of the QCD axion mass.
As a rule, it is (implicitly) assumed
that the axion has no mass at tree level, so that all of it 
is generated by the SU(3) gauge dynamics. Then, the mass 
correction can be evaluated at $\omega = 0$, in which 
case it is proportional to the so-called (Euclidean)
topological susceptibility.
Though the problem is technically 
challenging, results have become available 
(cf., e.g.,\ refs.~\cite{topo0,topoT} and references therein), 
and we return to them presently (cf.\ \eq\nr{chi_topo}).

However, a mass determined 
at $\omega = 0$ is correct only if there is no ``bare mass'' from
an UV theory. This is a problem, since 
our inflaton potential $V^{ }_0$ already contains a mass; 
it is envisaged to have been generated by an UV gauge theory, at
a scale higher than the IR one on which we focus. In this situation, 
the contribution to the mass through the 
SU(3) topological susceptibility is only a correction, and its
determination at $\omega = 0$ rather than $\omega\simeq m$ represents 
an uncontrolled approximation from the physics point of view.

Despite these reservations, 
let us estimate how large the mass correction could be. 
We denote by 
$\chi^{ }_\rmi{topo}$ the SU(3) topological susceptibility,
and by $t^{ }_0$ an auxiliary quantity often
used for setting the scale in lattice simulations.
Then, at $T=0$, 
$t_0^2\,  \chi^{ }_\rmi{topo} = 6.67(7) \times 10^{-4}_{ }$~\cite{topo0}, 
whereas examples of thermal values are 
$t_0^2\,  \chi^{ }_\rmi{topo} = 2.25(12) \times 10^{-5}_{ }$
at $T \sqrt{8 t^{ }_0}=1.081$
and
$t_0^2\,  \chi^{ }_\rmi{topo} = 3.43(27) \times 10^{-6}_{ }$
at $T \sqrt{8 t^{ }_0}=1.434$~\cite{topoT}. Inserting 
a conversion to the critical temperature, 
$
 \Tc^{ } \sqrt{t^{ }_0} = 0.2489(14)
$~\cite{Tc}, 
a rough qualitative represention, 
incorporating an expected
functional dependence at higher temperatures, is 
\be
 \chi^{ }_\rmi{topo} 
 \; \stackrel{ T \lsim 0.95 \Tc^{ } }{\simeq} \; 
 0.17 \, \Tc^4 
 \;, \qquad
 \chi^{ }_\rmi{topo}
 \; \stackrel{  T \gsim 0.95 \Tc^{ } }{\simeq} \;
 \frac{ 0.12 \, \Tc^{11} }{ T^7 } 
 \;, \la{chi_topo}
\ee 
and the corresponding mass correction from the IR gauge theory
evaluates to 
\be
 \delta m^2_\rmiii{IR}
 \big|^{ }_{\omega = 0}
 \;\simeq\;
 \frac{\chi^{ }_\rmii{topo}}{f_a^2}
 \;. \la{mass_formula}
\ee

Let us connect $ \delta m^2_\rmiii{IR} $ to the parameters that 
appear in $V^{ }_0$. We have referred to $V^{ }_0$ as the tree-level
potential, but let us now assume that it also includes those radiative
and thermal corrections which do not change the shape of $V^{ }_0$
(in contrast, shape-changing structures lead 
to what we have denoted by $V$, 
as discussed in \se\ref{ss:s_varphi}).
Let then $m_0^2$ be the value of the mass {\em before} the 
inclusion of the IR contribution. Then, at zero temperature, 
\be
 m^2_{ } |^{ }_{T = 0} =  m_0^2 + \delta m^2_\rmiii{IR} |^{ }_{T = 0}
 \;,
\ee
whereas a would-be thermal mass squared reads
\be
 m^2_{\T}  = m_0^2 + \delta m^2_{\rmiii{IR},\T} 
           = m^2_{ } |^{ }_{T = 0} 
                       - \delta m^2_\rmiii{IR} |^{ }_{T = 0}   
                       + \delta m^2_{\rmiii{IR},\T} 
 \;. \la{dmm}
\ee
According to \eq\nr{chi_topo}, 
$
                         \delta m^2_{\rmiii{IR},\T} 
                       < \delta m^2_\rmiii{IR} |^{ }_{T = 0}   
$, 
so the thermal correction in \eq\nr{dmm} is negative: 
the effective mass squared decreases at high temperatures. 

%
%

After these order-of-magnitude estimates, 
let us explain why the mass correction should be unimportant. 
First of all, inserting numerical values from \eq\nr{params}
and estimating the bare mass as 
$m^2 \sim \Lambda_\rmiii{UV}^4/f_a^2$, in accordance with 
\eq\nr{mass_formula}, the scale of the UV gauge theory is 
$\Lambda^{ }_\rmiii{UV} \sim \sqrt{m f^{ }_a} \sim 10^{-3}_{ }\mpl^{ }$. 
Now, according to \se\ref{ss:Tmax}, the solutions fall in two 
different classes. If 
$\Tmax^{ } \ge  \Tc^{ }$, then 
$
  | \delta m^2_{\rmiii{IR},\T} - \delta m^2_\rmiii{IR} | ^{ }_{T=0} | \sim
  \LambdaIR^4 / f_a^2 \ll 
  \LambdaUV^4 / f_a^2 \sim m^2
$,
i.e.\ the thermal mass correction is exceedingly small. 
If $\Tmax^{ } < \Tc^{ }$, 
the system stays in the confined phase, 
and there is no thermal mass correction. 

It is for these reasons, as well as the conceptual issues explained
at the beginning of this section, that we omit thermal mass 
corrections in the following. 

%
\subsection{Inflaton potential and parameter choices}
\la{ss:params}

In order to study heating-up in a semi-realistic framework, 
we consider axion-like (or natural) inflation~\cite{ai}, 
and fix the parameters of the potential 
to agree with Planck data~\cite{planck}. 
For this purpose, we adopt the predictions of cold inflation, despite 
the fact that the solution is technically unstable 
(cf.\ \se\ref{ss:upsilon}). The coupling $\alpha$
does not appear in these predictions, 
and can thus be freely varied for understanding
heating-up dynamics. In principle
it would be interesting to verify 
{\it a posteriori} how significantly phenomenological
predictions are altered by thermal effects if high
temperatures are reached already during the inflationary stage
(for a review of warm inflation see, e.g., ref.~\cite{warm_review}). 
However, this dynamics is physically distinct from and takes place
much earlier than the heating-up stage that we are interested in.

For the inflaton potential, we take the ansatz 
\be
 V^{ }_0
 \;\simeq\;
 m^2 f_a^2\, 
   \biggl[ 1 - \cos\biggl( \frac{\bar\varphi}{f^{ }_a} \biggr) \biggr]
 \;, \quad
 V^{ }_{0,\varphi}
 \;\simeq\;
 m^2 f_a^{ }\, 
    \sin\biggl( \frac{\bar\varphi}{f^{ }_a} \biggr)
 \;,
 \la{V}
\ee
whereas 
$
 V^{ }_{0,\T}
$
is approximated as small
(cf.\ \se\ref{ss:mass}).
In the numerical estimates the parameters
are fixed to benchmark values from ref.~\cite{scan}, 
\be 
 f^{ }_a = 1.25 \, \mpl^{ } 
 \;, \quad 
 m = 1.09 \times 10^{-6}_{ }\, \mpl^{ } 
 \;, \quad
 \bar\varphi(t^{ }_\rmi{ref}) = 3.5 \,\mpl^{ } 
 \;, \la{params}
\ee
where 
$
 t^{ }_\rmi{ref}
$
denotes the time at which we start the simulation, 
conveniently chosen as 
\be
  t^{ }_\rmi{ref} 
  \; \equiv \; 
  \sqrt{\frac{3}{4\pi}} 
  \frac{ m^{ }_\rmiii{pl} }{m \bar\varphi(t^{ }_\rmiii{ref})}
 \;. \la{tref}
\ee
This is within $\rmO(1)$ of the initial inverse Hubble rate.  

%
\subsection{When is the thermalization assumption self-consistent?}
\la{ss:equil}

To conclude this section, let us return to the important question
of when the temperature is a useful concept. 
An upper bound extending almost up to the 
Planck scale was presented in \se\ref{se:intro}, 
however it was based on the assumption that the Hubble rate is already
dominated by the energy density carried by radiation. 

The question of how fast a general system equilibrates is
a hard one. Frequently, the dynamics following inflation is studied
by solving classical field equations of motion.\footnote{%
 This method is referred to as 
 preheating~\cite{preheat1,preheat2}.
} 
The problem is that classical field dynamics can be
correct only for large occupation numbers, not in the typical 
domain where the occupation is of order unity. But it is precisely
momenta from the latter domain, $p\sim \pi T$, which carry most of
the radiation energy density. In other words, the issue of thermalization 
cannot be properly resolved with classical field theory. 
In the heavy-ion context, where thermalization of non-Abelian systems
has been studied extensively, 
the method of choice relies nowadays rather
on effective kinetic theory~\cite{equil}. 

\begin{figure}[t]

\begin{eqnarray*}
  && \hspace*{-2.5cm}
 \AmplGauge  \hspace*{+3.0cm} 
 \AmplInfla  \hspace*{-1.0cm} 
  \\
  && \hspace*{-0.85cm} \mbox{(a)}
     \hspace*{+5.1cm} \mbox{(b)}
\end{eqnarray*}

\vspace*{-3mm}

\caption[a]{\small
  (a) 
  example of an elastic process responsible for
  the thermalization of gauge fields (wiggly lines); 
  (b)
  an artistic impression of 
  an inelastic process contributing to the thermalization 
  of the inflaton (dashed line). Given the peculiar nature 
  of the operator in \eq\nr{L}, indicated by the blob, 
  the gauge configurations
  are non-perturbative here (if $\omega \ll m$). 
  Therefore we have sketched them with multiple wiggly lines.
}

\la{fig:equil}
\end{figure}

If we do think in the language of effective kinetic theory, 
we can draw diagrams responsible for thermalization. Examples
for a non-Abelian plasma and for the inflaton, respectively, 
are illustrated in \fig\ref{fig:equil}. 
The gauge plasma equilibration rate 
(i.e.\ the thermally averaged amplitude squared)
is then $\Gamma^{ }_g \sim \alpha^2 T$
within the weak-coupling expansion, 
whereas the inflaton equivalent is 
$\Gamma^{ }_\varphi \sim \Upsilon^{ }_\rmiii{IR}$.
In the subsequent sections, we will compare these with
the Hubble rate {\it a posteriori}, while the computations
themselves are carried out in the presence 
of a temperature-like parameter, 
reminiscently of the scenario of warm inflation~\cite{warm_review}.
Here it is appropriate to remark that the general objections to 
warm inflation, raised in ref.~\cite{linde}, are avoided in 
the non-Abelian axion inflation context, as it is
possible to have 
a large thermal friction coefficient (cf.\ \se\ref{ss:upsilon})
without inducing a large thermal mass correction (cf.\ \se\ref{ss:mass}).

%
\section{Temperature evolution}
\la{se:T}

Before attacking the full numerical solution
of \eqs\nr{eom_field}, \nr{eom_plasma}, 
and \nr{eom_plasma_2} (cf.\ \se\ref{ss:Tmax}), 
we introduce
the concept of a stationary temperature, $\Tstat^{ }$, 
which already helps us 
to understand the parametric dependence 
of the temperature scale reached (cf.\ \se\ref{ss:Tstat}).

%
\subsection{Stationary temperature as a qualitative estimate}
\la{ss:Tstat}

The existence of a stationary temperature at intermediate
stages of warm natural inflation follows from 
the argumentation in refs.~\cite{warm_axion1,fixed_pt}. 
Physically, 
this corresponds to a situation in which the 
energy released from the inflaton to radiation through friction, 
precisely balances against the energy
diluted by the Hubble expansion. 
After a while, 
$T$ starts to increase above $\Tstat^{ }$,
obtaining a maximal value, $\Tmax^{ }$. 
While estimating $\Tmax^{ }$
requires the solution of a coupled set of differential
equations, it is easier to determine
$\Tstat^{ }$, as the equations are algebraic, 
and we may furthermore employ slow-roll approximations in them. 
Yet, as consolidated by the numerical studies in \se\ref{ss:Tmax},  
$\Tstat^{ }$ already gives
an order-of-magnitude estimate of $\Tmax^{ }$. 

Let us consider the solution of \eqs\nr{eom_field} and \nr{eom_plasma}
in the slow-roll regime. Then \eq\nr{eom_field} implies that
\be
 \dot{\bar\varphi} \; \simeq \; - \frac{V^{ }_\varphi}{ 3 H + \Upsilon}
 \;. \la{varphi_slowroll}
\ee
In \eq\nr{eom_plasma}, 
we search for a stationary solution, 
with $\dot{e}^{ }_r - T \dot{V}^{ }_\T \simeq 0$.
Recalling the thermodynamic relation 
$e^{ } + p^{ } = T s^{ }$, where
$s^{ }$ is the entropy density, yields the master relation
\be
 3 \Tstat^{ } s 
 \; 
 \stackrel{ 
          }{\simeq} 
 \;
 \frac{\Upsilon}{H} \frac{V_\varphi^2}{(3 H + \Upsilon)^2}
 \;. \la{master}
\ee
As further simplifications,
the Hubble rate can be approximated as 
$
 H \simeq \sqrt{8\pi V/(3\mpl^2)}
$
during the slow-roll period, 
and we may furthermore set 
$
 \bar\varphi \to 
 \bar\varphi(t^{ }_\rmi{ref})
$
in 
$V$ and $V^{ }_\varphi$. 

\begin{figure}[t]

\hspace*{-0.1cm}
\centerline{%
   \epsfxsize=7.5cm\epsfbox{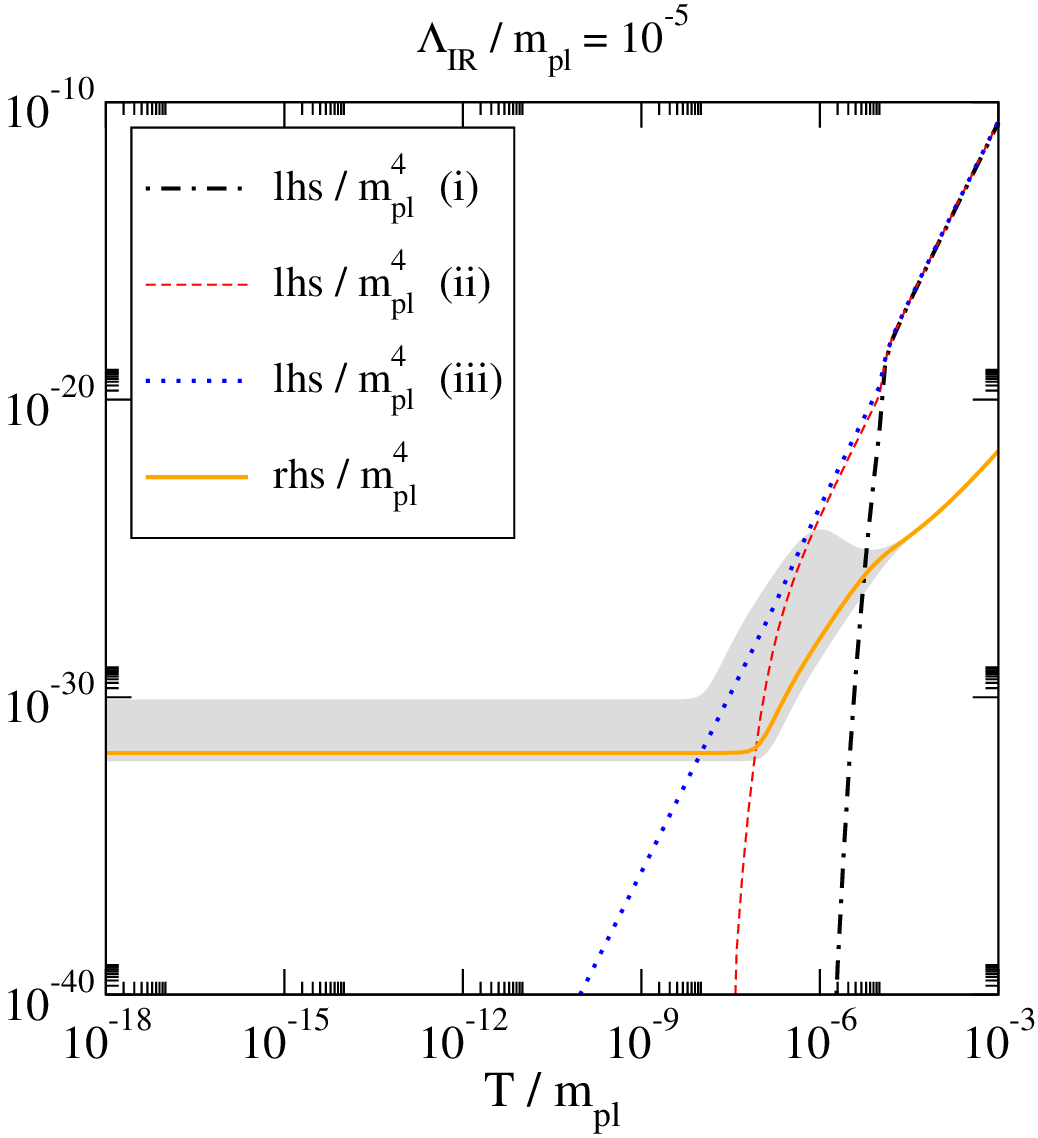}
   ~~~\epsfxsize=7.5cm\epsfbox{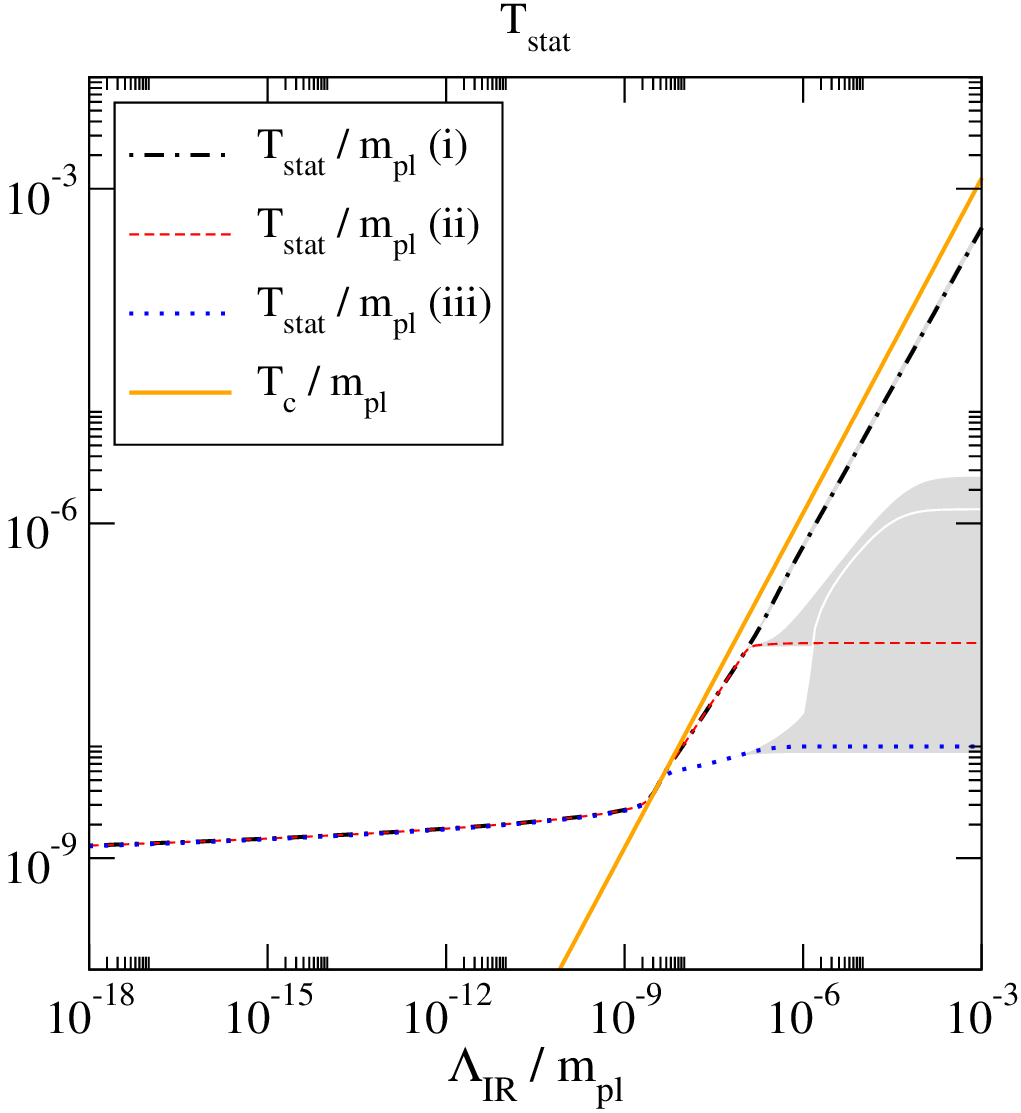}
}

\caption[a]{\small
  Left: the left-hand side (lhs) and right-hand side (rhs) of 
  \eq\nr{master}, in units of $\mpl^4$, with the former
  evaluated for the three cases 
  defined in \se\ref{ss:Tstat}.
  In this example the scale
  parameter has been set to $\LambdaIR^{ } = 10^{-5}_{ }\mpl^{ }$.
  The grey band corresponds to the variation $x\in(0.2,2.0)$ 
  in \eq\nr{alpha}. 
  Right: the solution (crossing point), denoted by $T^{ }_\rmii{stat}$, 
  as a function of $\LambdaIR^{ } / \mpl^{ }$. For comparison 
  we also show the critical temperature $\Tc^{ }$.
}

\la{fig:Tstat}
\end{figure}

We illustrate the solution originating from 
\eq\nr{master} in three qualitatively different cases: 
\bi

\item[(i)]
{\bf confining plasma, non-thermal inflaton.}
In this case we insert the equation of state from \se\ref{ss:s_r},
$s \to s^{ }_r$,
and assume that the inflaton does not thermalize.
This assumption can often be justified 
{\it a posteriori}, cf.\ \se\ref{ss:Tmax}.

\item[(ii)]
{\bf confining plasma, thermalized inflaton.}
We add the contribution of a thermalized inflaton 
to the entropy density, 
$ s \to s^{ }_r - V^{ }_\T$,
as specified in \se\ref{ss:s_varphi}.

\item[(iii)]
{\bf confining plasma with a light degree of freedom and thermalized inflaton.}
We further add
one free massless bosonic 
degree of freedom to the radiation plasma, in order
to illustrate the qualitative
influence of a dark photon or a dark light pion. 
Then 
$
 s \to s^{ }_r - V^{ }_\T + 2\pi^2 \Tstat^3/45
$ in \eq\nr{master}. 

\ei

For the three cases defined, 
the left and right-hand sides (lhs, rhs)
of \eq\nr{master}
are illustrated in \fig\ref{fig:Tstat}(left). The resulting values of 
$\Tstat^{ }/\mpl^{ }$, from the crossings of the respective 
curves, are plotted in \fig\ref{fig:Tstat}(right). 

It can be observed from \fig\ref{fig:Tstat}(right) that there is 
a specific domain of $\LambdaIR^{ }$ at which the behaviour of 
the system changes. 
Let us denote the smallest value of $\LambdaIR^{ }$ for which 
the system stays stationary at exactly the critical temperature  
by $[\,\Lambdaref^{ }\,]^{ }_\rmi{stat}$. 
This corresponds to 
\be
 3 \Tc^{ }\, s
 \; 
 \underset{ \LambdaIR^{ } \to [\Lambdaref^{ }]^{ }_\rmii{stat} }{  
  \overset{ T \to \Tc^{ } }{ \equiv } } 
 \;
 \frac{\Upsilon}{H} \frac{V_\varphi^2}{(3 H + \Upsilon)^2}
 \quad \Rightarrow \quad 
 \bigl[\, \Lambdaref^{ } \bigr]^{ }_\rmi{stat}
 \; \simeq \;  
 3\times 10^{-9}_{ }\, \mpl^{ } 
 \;. \la{Lambda0}
\ee
A similar logic will be applied in \se\ref{ss:Tmax}, with however
the stationary temperature replaced by the maximal one, 
yielding then an outcome denoted by $\Lambdaref^{ }$.
In between $ [\,\Lambdaref^{ } \,]^{ }_\rmi{stat}$ and 
$\Lambdaref^{ }$, the system heats up to above $\Tc^{ }$, and 
experiences two nearby phase transitions, 
before and after this moment. 
At $\LambdaIR^{ } > \Lambdaref^{ }$,  
$\Tmax^{ } < \Tc^{ }$, and no phase transition takes place.  

Now, in terms of \fig\ref{fig:Tstat}(left), 
we find that when $\LambdaIR^{ } = [\,\Lambdaref\,]^{ }_\rmi{stat}$, 
then the lines cross in a domain where the rhs curve is flat. 
The cusp in the rhs curve is where the behaviour of $\Upsilon$
changes, from $\Upsilon^{ }_\rmiii{UV}$ at low temperatures 
to $\Upsilon^{ }_\rmiii{IR}$ at high temperatures. Therefore, 
$[\Lambdaref^{ }]^{ }_\rmi{stat}$ 
is determined by $\Upsilon^{ }_\rmiii{UV}$, 
and independent 
of the non-perturbative physics of $\Upsilon^{ }_\rmiii{IR}$
that originates from sphaleron dynamics. 
The latter is important in the regime 
$
 \LambdaIR^{ } \gg [\Lambdaref^{ }]^{ }_\rmi{stat}
$.

%
\subsection{Maximal temperature from a numerical solution}
\la{ss:Tmax}

\begin{figure}[t]

\hspace*{-0.1cm}
\centerline{%
   \epsfysize=5.0cm\epsfbox{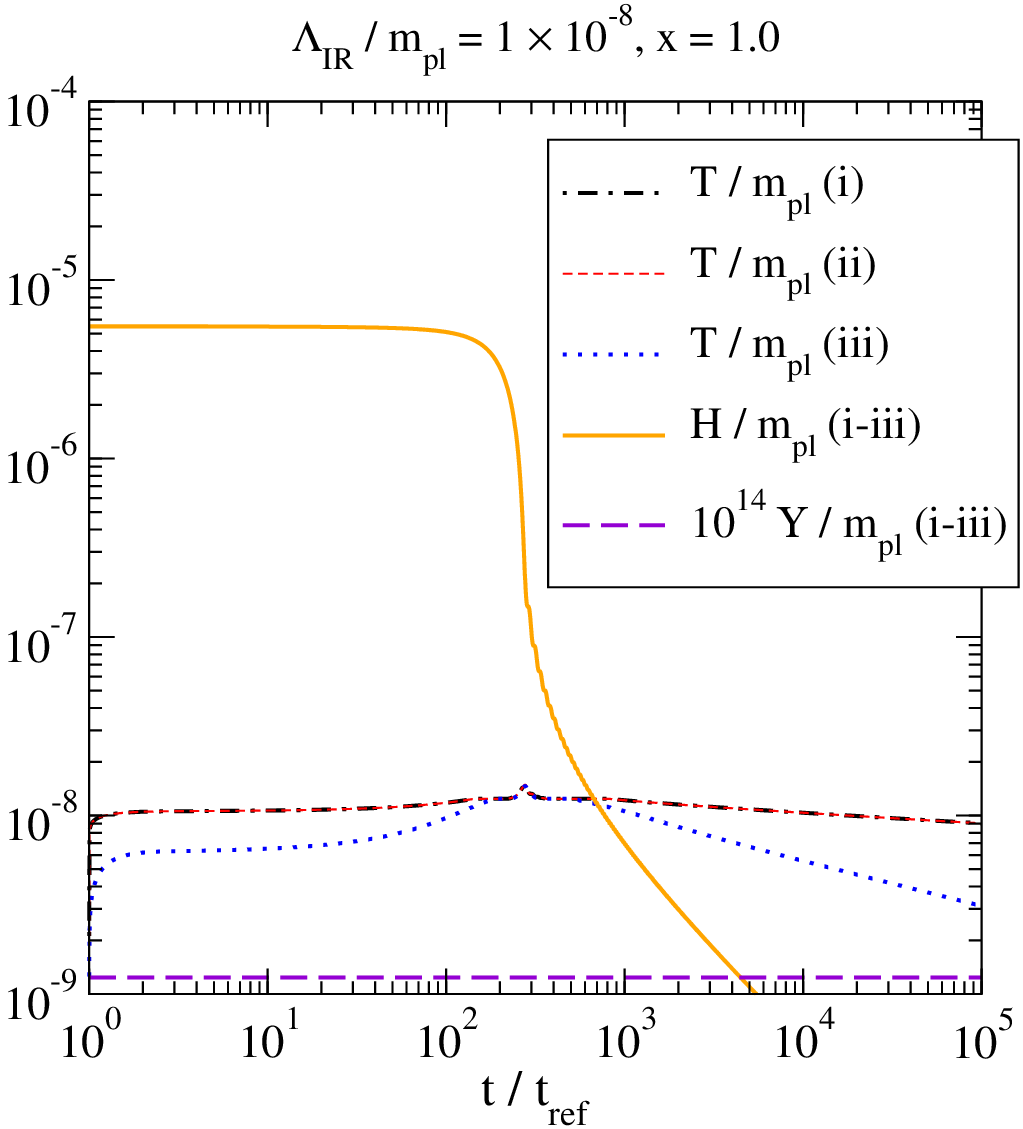}
   ~~~\epsfysize=5.0cm\epsfbox{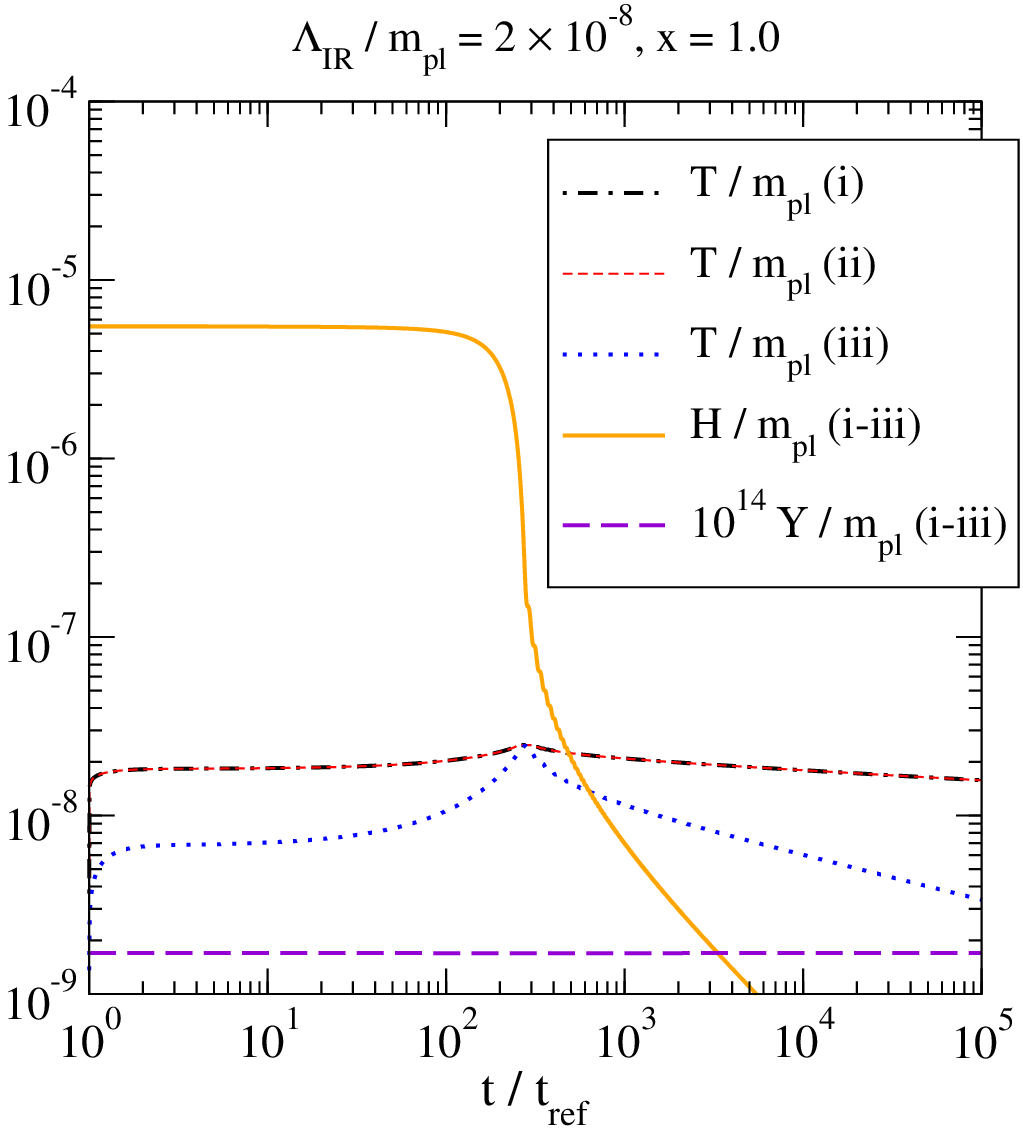}
   ~~~\epsfysize=5.0cm\epsfbox{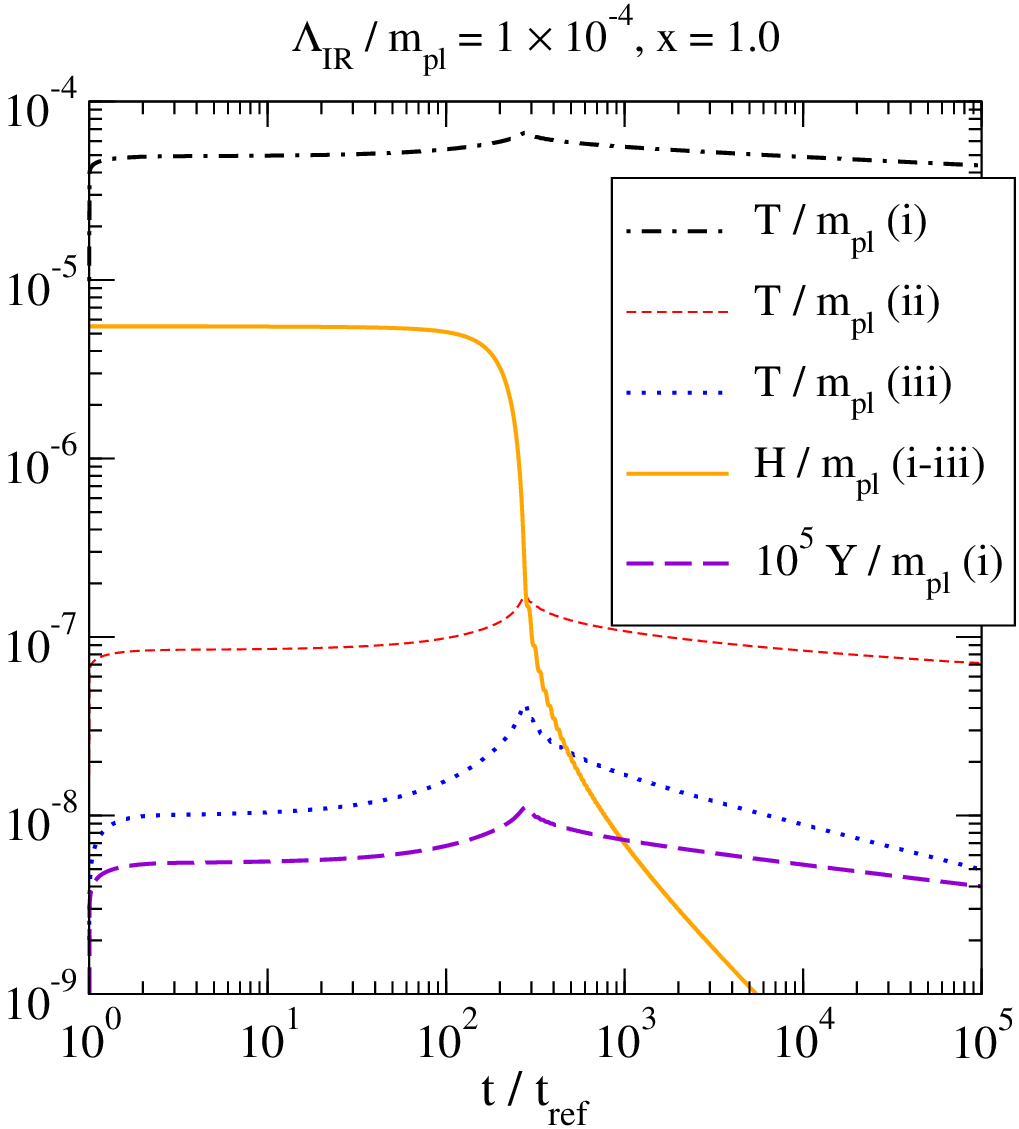}
}

\caption[a]{\small
  Examples of solutions in which $\Tc^{ }$ is crossed (left), 
  reached (middle), and not reached (right). 
  For the first two cases, the corresponding 
  volume fractions are shown in \fig\ref{fig:mixed}.
  In the right-most panel, 
  the temperature of case~(i) far exceeds the Hubble rate, 
  but the friction coefficient $\Upsilon$ remains small 
  (the implications from here are discussed in the text).
}

\la{fig:soln}
\end{figure}

\begin{figure}[t]

\hspace*{-0.1cm}
\centerline{%
   ~~~\epsfxsize=7.5cm\epsfbox{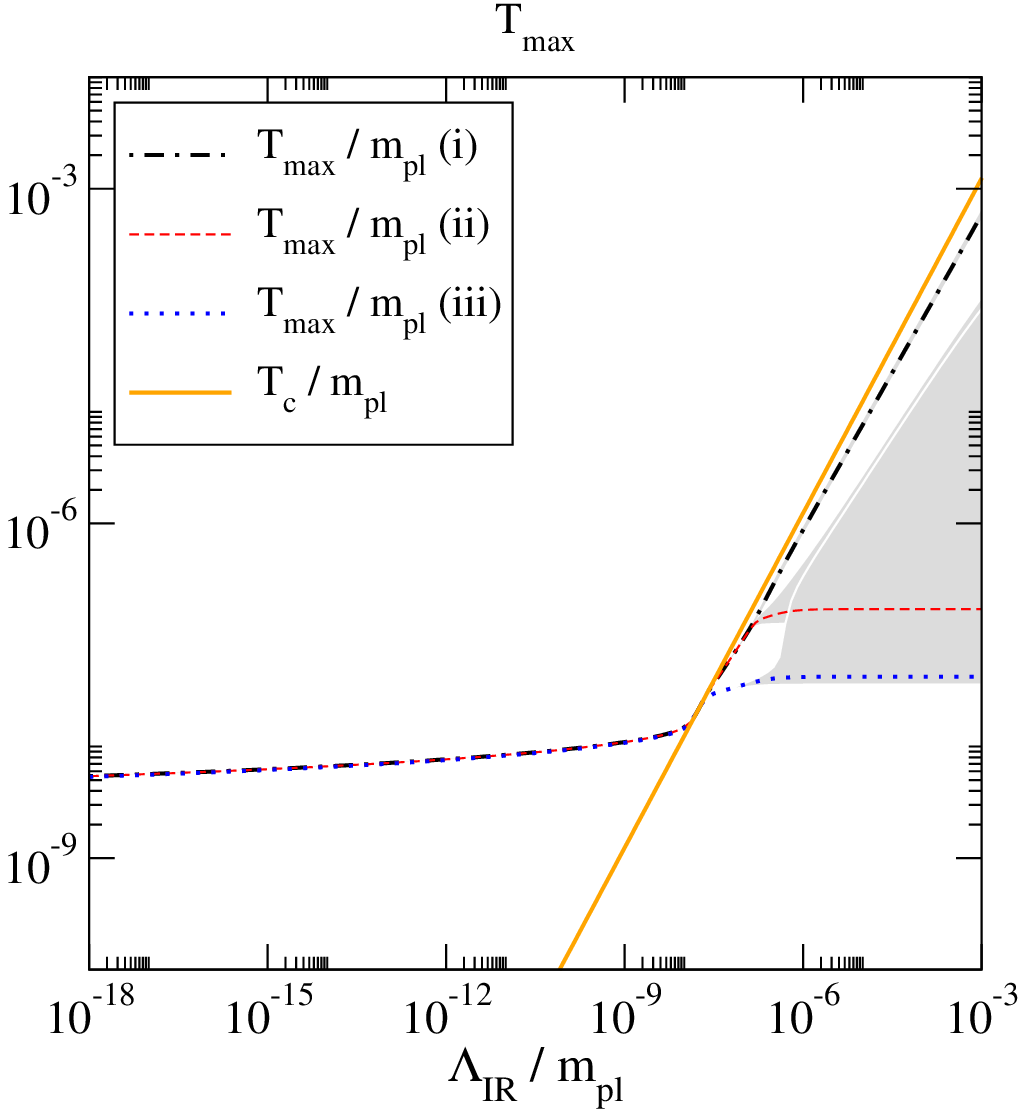}
}

\caption[a]{\small
  Scan of $\Tmax^{ }$, for the three prototype
  systems defined in \se\ref{ss:Tstat}, 
  as a function of $\LambdaIR^{ } / \mpl^{ }$. 
  Our setup is self-consistent only for 
  $\LambdaIR^{ } < \LambdaUV^{ }\sim 10^{-3}\mpl^{ }$, 
  so we restrict the axis to this domain. 
  The grey bands show 
  the effect of varying the parameter $x$ in \eq\nr{alpha} in 
  the range $x \in (0.2,2.0)$.
  The uncertainties are huge for cases (ii) and (iii) 
  at large $\LambdaIR^{ }$, as we then need to evaluate 
  the friction coefficient $\Upsilon$ deep in the confined phase. 
  By $\Tc^{ }$ we denote the critical temperature. 
}

\la{fig:Tmax}
\end{figure}

We now move on to consolidate the qualitative results
from \se\ref{ss:Tstat} through a full solution of 
\eqs\nr{eom_field}, \nr{eom_plasma}, and \nr{eom_plasma_2}.
For this we 
consider the same three cases as defined in \se\ref{ss:Tstat}
and illustrated in \fig\ref{fig:Tstat}.
As initial conditions, we take 
$T(t^{ }_\rmi{ref}) \simeq 0.2 \Tstat^{ }$, 
$\bar\varphi(t^{ }_\rmi{ref})$ from \eq\nr{params}, 
and the slow-roll evolution rate 
$\dot{\bar\varphi}(t^{ }_\rmi{ref}) \simeq - V^{ }_\varphi / (3 H)$. 
However, the solution is an attractor, and therefore soon 
independent of the initial conditions. The domain in which
the system heats up exactly to $\Tc^{ }$ is denoted by 
$\Lambdaref^{ }$, with the numerical value
$\Lambdaref^{ }\sim 2\times 10^{-8}_{ }\mpl^{ }$.

Examples of solutions are shown in \fig\ref{fig:soln}, and a scan
of the resulting values of $\Tmax^{ }$ in \fig\ref{fig:Tmax}.
We conclude that 
\bi

\item
if $\LambdaIR^{ } < \Lambdaref^{ }$, $\Tmax^{ } > \Tc^{ }$, 
and the system undergoes a phase transition as it cools down. 
For $[\Lambdaref^{ }]^{ }_\rmi{stat} < \LambdaIR^{ } < \Lambdaref^{ }$, 
it also undergoes a nearby previous transition as it heats up, 
as visible in \fig\ref{fig:soln}(left). 
However, 
in this domain
$
 \alpha^2_{ }T \ll T \ll H
$, 
so it is questionable whether the
temperature has a literal meaning
during the first transition (cf.\ \se\ref{ss:equil}). 
The maximal temperature has the numerical value 
$
 \Tmax^{ } \sim \Lambdaref^{ } \sim 2\times 10^{-8}_{ }\mpl   
$.

\item
if $\LambdaIR^{ } \sim \Lambdaref^{ }$, 
then $\Tmax^{ } = \Tc^{ }$:
the system heats up to $\Tc^{ }$ 
(cf.\ \figs\ref{fig:mixed}(right) and \ref{fig:soln}(middle)). 

\item
if $\LambdaIR^{ } > \Lambdaref^{ }$, $\Tmax^{ } < \Tc^{ }$, 
and the system undergoes no phase transition. 
Nevertheless it heats up to 
a high temperature, $\Tmax^{ }\sim \LambdaIR^{ }$,  
if the plasma is confining. If the inflaton
thermalizes, the maximal temperature is cut off by 
the inflaton mass, $\Tmax^{ }\sim \min(m,\LambdaIR^{ })$. 
However, it appears unlikely that the inflaton thermalizes, 
because its would-be thermalization rate 
$\sim \Upsilon^{ }_\rmiii{IR}$ is much
below the Hubble rate, cf.\ \fig\ref{fig:soln}(right).
If the plasma includes $g^{ }_*$ massless degrees of freedom, 
the maximal temperature stays 
at~$\Tmax^{ }\sim \Lambdaref^{ }(2\dA /g^{ }_*)^{1/4}_{ }$, 
irrespective of the value of $\LambdaIR^{ }$.
However, this consideration assumes that the massless 
degrees of freedom thermalize, i.e.\ that their full phase
space can be filled to accommodate the entropy released
from inflaton oscillations.\footnote{%
 For the example of dark pions, the equilibration
 rate from $2\to 2$ scatterings is $\Gamma^{ }_\pi \sim T^5/f_\pi^4$, 
 where $f^{ }_\pi \sim \LambdaIR^{ }\sim \Tc$. 
 If we assumed thermalization
 and found $\Tmax^{ } \sim \Tc^{ }$, 
 the assumption would be self-consistent. 
 However, since the thermalization assumption leads 
 generically to $\Tmax^{ }\ll \Tc^{ }$, so that 
 $\Gamma^{ }_\pi \ll \Gamma^{ }_g$, it is less so. 
} 

\ei

It can be observed from \fig\ref{fig:Tmax} that in cases
(ii) and (iii), there is a large uncertainty in the value of 
$\Tmax^{ }$. This originates from the gauge coupling $\alpha$, through 
the parameter $x$ (cf.\ \eq\nr{alpha}). 
The reason is that these would-be solutions
lie deep in the confined phase, where our
treatment of $\Upsilon$ is not reliable. However, as discussed
above, the physical significance of these solutions is questionable
for another reason as well, namely that the thermalization assumption 
is hard to consolidate. It is a lucky coincidence that both the 
technical and conceptual uncertainties can be 
reflected by the same error bands. 

%
\section{Physical implications for gravitational waves}
\la{se:phys}

The patterns observed in \se\ref{ss:Tmax} have a number of 
potential implications for primordial gravitational waves, 
to which we now turn.

%
\subsection{Total energy density in the gravitational wave background}
\la{ss:gw_neff}

A thermal plasma necessarily generates a spectrum of 
gravitational waves, whose energy density 
contributes to the effective number of massless
degrees of freedom, $N^{ }_\rmi{eff}$~\cite{gravity_qualitative}.
As long as the maximal temperature 
is below $\sim 10^{-2}_{ }\mpl^{ }$, 
the contribution to $N^{ }_\rmi{eff}$ 
is small~\cite{gravity_lo,srt,gravity,rt,gs}. 
At the same time, 
our framework is consistent as long as 
$\LambdaIR^{ } < \LambdaUV^{ }\sim 10^{-3}_{ }\mpl^{ }$.
Given that $\Tmax^{ } < \LambdaIR^{ }$ for large values
of $\LambdaIR^{ }$, we then also have
$\Tmax^{ } < \LambdaUV^{ }$. 
Therefore none of the heating-up scenaria
that we have found is excluded by 
constraints from $N^{ }_\rmi{eff}$. 

%
\subsection{Monotonically growing spectrum from the hottest epoch}
\la{ss:gw_f03}  

Thermal fluctuations generate a monotonically
increasing intermediate-frequency component 
in the gravitational wave background~\cite{scan}, 
originating dominantly from when $T\sim \Tmax^{ }$,
\be
 \Omega^{ }_\rmii{GW} h^2 
  \;\;\stackrel{10^{-6}_{ }\,{\rm Hz}\,
                   \le \, \fnow^{ }  \, \le \,
              10^{2}_{ }\,{\rm Hz}\;\; }{\supset}\; 
  A\, 
  \biggl( \frac{\fnow^{ }}{\mbox{Hz}} \biggr)^3
 \biggl( \frac{T\eta}{\mpl^4} \biggr)^{ }_\rmi{max}
 \;, \quad 
 \bigl( T\eta \bigr)^{ }_\rmi{max} 
 \sim \frac{\Tmax^4}{\alpha_\rmi{min}^2}
 \;, \la{scan}
\ee
where $\eta$ denotes the shear viscosity and the parametric behaviour
shown applies within the weak-coupling expansion. 
Close to the upper bound 
$\Tmax^{ }\sim \LambdaUV^{ } \sim 10^{-3}_{ }\mpl^{ }$, 
this background could become marginally observable 
at the highest frequencies $\fnow^{ }\sim 100$~Hz
that are perhaps probed in the future by 
the Einstein telescope
and 
the DECIGO interferometer.
It would be interesting to carry out a more quantitative
sensitivity study of this possibility.\footnote{%
 We thank Germano Nardini for drawing our attention to this prospect.
 As far as the numerical coefficient in \eq\nr{scan} goes,  
 ref.~\cite{scan} estimated  
 $A \sim 10^{-9}_{ }$ for $\LambdaIR^{ }\sim 10^{-20}_{ }\mpl^{ }$, 
 but the value could be different 
 at larger~$\LambdaIR^{ }$, 
 given that it depends on the cosmological expansion history.
 }

%
\subsection{Peaked spectra from first-order phase transitions}
\la{ss:gw_transition}

If $\LambdaIR^{ } < \Lambdaref^{ }$, we find that 
$\Tmax^{ } > \Tc^{ }$. As the system cools down, it then 
undergoes a thermal first-order phase transition, which
may lead to a gravitational wave signal~\cite{ps}.
In our particular example, the transition is a weak one, 
with $\Delta e(\Tc^{ }) / e(\Tc^{ }) \ll 1$, but it could
conceivably be stronger in other strongly coupled theories. 

In order to discuss the significance of such transitions, 
let us envisage causal bubble dynamics 
taking place with a characteristic length scale 
$\ell^{ }_\rmii{B} \ll \ell^{ }_\rmii{H} \equiv H^{-1}_{ }$.
Three different temperatures play a role, 
the critical temperature ($\Tc^{ }$); 
the temperature at which radiation takes over from 
$\bar\varphi$ as the dominant component 
of the energy density ($\Te^{ }$); 
and the temperature today ($\Tnow^{ }$). 
Within our computation, $\Te^{ }$ is reached approximately
when $H \simeq \Upsilon$, but it might be reached sooner
if preheating dynamics were accounted for~\cite{preheat1,preheat2}. 
We may redshift $\ell^{ }_\rmii{B}$ as
$
 \ell^{ }_\rmii{B}(\Tnow^{ }) = 
 \frac{a(\Tnow^{ })}{a(\Te^{ })}\, \frac{a(\Te^{ })}{a(\Tc^{ })}\,
 \ell^{ }_\rmii{B}(\Tc^{ })
$.
The current-day frequency corresponding to this wavelength
is $\fnow^{ } \simeq c /\ell^{ }_\rmii{B}(\Tnow^{ })$. 
Expressing the first ratio of the scale factors with the help of 
entropy densities, and the second with $e$-folds, 
yields (in natural units)
\be
 \frac{\fnow^{ }}{\mbox{Hz}}
 \; \simeq \; 
 \left\{
 \begin{array}{ll} 
  \displaystyle
  \mbox{s}\, \Tnow^{ } \, 
  \biggl( \frac{\snow^{ } / \Tnow^3}
               {s^{ }_{\rm e} / \Te^3} \biggr)^{\fr13}_{ } \,
  e^{ -\Delta N^{ }_{{\rm c}\to{\rm e}} }_{ }
  \frac{\ell^{ }_\rmii{H}}{\ell^{ }_\rmii{B}} \, 
  \frac{H(\Tc^{ })}{\Te^{ }}
  & 
  \;, \quad \Tc^{ } > \Te^{ }
  \\
  \displaystyle
  \mbox{s}\, \Tnow^{ } \, 
  \biggl( \frac{\snow^{ } / \Tnow^3}{s^{ }_{\rm c} / \Tc^3}
  \biggr)^{\fr13}_{ } \,
  \frac{\ell^{ }_\rmii{H}}{\ell^{ }_\rmii{B}} \, 
  \frac{H(\Tc^{ })}{\Tc^{ }}
  & 
  \;, \quad \Tc^{ } < \Te^{ } 
  \end{array}
 \right.
 \;. \la{f0}
\ee
For $\LambdaIR^{ } \simeq \Lambdaref^{ }$ we find 
$\Tc^{ } \simeq 2\times 10^{-8}_{ }\,\mpl^{ }$; 
$\Te^\rmi{(i)} \simeq 3\times 10^{-9}_{ }\,\mpl^{ }$  and 
$\Te^\rmi{(iii)} \simeq 2\times 10^{-12}_{ }\,\mpl^{ }$, where the 
superscript refers to the case defined in \se\ref{ss:Tstat}; 
and 
$\Delta N^{ }_{{\rm c}\to{\rm e}} \simeq 23$. 
For $\LambdaIR^{ } < \Lambdaref^{ }$, 
$\Tc^{ }$ decreases in proportion to $\LambdaIR^{ }$, 
and goes ultimately below~$\Te^{ }$.
Inserting 
$ \Tnow^{ } \approx 2.7255 \mbox{\hspace*{0.4mm}K} $, 
$\ell^{ }_\rmii{H} / \ell^{ }_\rmii{B} 
\simeq 10^{2 ... 4}_{ }$,\footnote{%
 In the gravitational wave literature,
 $\ell^{ }_\rmii{H} / \ell^{ }_\rmii{B}$ is frequently
 denoted by $\beta / H$ (up to a factor of velocity). 
 }
and assuming that at $T \le \Te^{ }$
the visible sector temperature is similar 
to the dark sector one, 
whereby Standard Model values can be adopted 
for the entropy densities~\cite{eos15}, gives 
$
 \fnow^\rmi{(i)} \lsim 5\times 10^{3 ... 5}_{ }\,
$Hz
and
$
 \fnow^\rmi{(iii)} \lsim 8\times 10^{6 ... 8}_{ }\,
$Hz.
A part of this range can be probed by interferometers, 
like again the Einstein telescope and ultimately perhaps DECIGO. 

In \figs\ref{fig:mixed}(left,middle), 
we also see another phase transition, 
passed as the system heats up. 
Concerning its significance, 
two issues should be raised. 
The first is that as the gravitational energy density
scales as $\sim 1/a^4$, the signal from the 
first transition is diluted by a factor
$
 \sim e_{ }^{-4 \Delta N}
$
compared with the second one, where $\Delta N$ is the number 
of $e$-folds between the transitions. 
If $\Delta N \gg 1$, then 
the signal gets diluted away. Only for a fine-tuned
value $\LambdaIR^{ }\sim 1.5\times 10^{-8}_{ }\mpl^{ }$, when the 
transitions are immediately adjacent to each other, could
the dilution be less spectacular. Second, as visible
in \figs\ref{fig:soln}(left,middle), the Hubble rate far exceeds
the temperature during the first transition, and also during
the second one if the two transitions are nearby. As the thermalization
rate is $\sim \alpha^2 T \ll H$ (cf.\ \se\ref{ss:equil}), 
it is questionable whether the temperature is a physically
meaningful notion in this case. Despite these concerns, 
it could be that non-equilibrium fluctuations 
produced during the first epoch could have a physical
effect, for instance by serving as nucleation seeds which
would permit for the second transition to proceed 
in a non-standard manner (cf.,\ e.g.,\ ref.~\cite{jdv}).  

%
\section{Summary and outlook}
\la{se:concl}

The purpose of this study has been to estimate the maximal 
temperature that strongly coupled dark sectors may reach. 
The basic point is that in the confined phase of such theories, 
thermodynamic functions such as the entropy density and the heat
capacity are exponentially small. Therefore, even a small release
of energy density from the inflaton field can heat up the system by
a large amount. For the very largest confinement scales,  
$\LambdaIR^{ } \sim (10^{-8}_{ } - 10^{-3}_{ })\mpl^{ }$, 
we find that the system heats up to close to the critical 
point, even if it remains just slightly below~$\Tc^{ }$
(cf.\ \fig\ref{fig:Tmax}). 
Interestingly, this most interesting scenario is treated
most reliably by our methods, as the gauge field thermalization 
rate clearly exceeds the Hubble rate, so that there is no doubt
about the validity of temperature as a physical notion
(cf.\ \fig\ref{fig:soln}(right)). 

As a particular consequence of such dynamics,
we have considered the gravitational wave background 
produced by thermal fluctuations around the heating-up epoch. 
In the frequency window
$(10^{-4}_{ } -10^{2}_{ })$~Hz, relevant for the LISA, 
Einstein telescope, and 
DECIGO interferometers, 
we predict a background increasing 
monotonically as $\sim \fnow^3$, with a coefficient 
proportional to the maximal shear viscosity of the plasma phase
(cf.\ \eq\nr{scan}). 
At the highest frequencies, 
the signal could be marginally observable 
in the future, though quantitative 
sensitivity studies would be needed for 
confirming or refuting this prospect. 

A complementary consequence is reached if the confinement 
scale is lowered, $\LambdaIR^{ } <  10^{-8}_{ }\mpl^{ }$. 
Then the system heats up above the critical 
temperature, confirming the possibility of a first-order
phase transition in a dark sector~\cite{ps}. 

We should underline that we have on purpose kept 
our study on a rather general level, with the hope that it is
then also more broadly applicable. Adding specific 
model assumptions, further issues could be addressed.
Notably, the value of the maximal temperature, and in particular
whether it is above $\Tc^{ }$ or not, 
has implications for dark matter production,
but the details are very
model-dependent~(cf., e.g., refs.~\cite{ulb,soni,dark}).

Apart from phenomenological issues, there are
also theoretical ingredients that could be built 
into our framework. Hoping that exploratory low-temperature 
investigations of the friction coefficient~\cite{explore,sphaleron_4d} 
turn ultimately into a semi-quantitative tool, the error bands
in \fig\ref{fig:Tmax} could be reduced. Were the same to happen
with the shear viscosity $\eta$, the gravitational wave estimate
in \se\ref{ss:gw_f03} could be sharpened.  
Inserting assumptions 
about the couplings of the inflaton to both the dark and the 
visible sector, and of the sectors between each other, 
two different temperatures could be tracked. Finally, 
investigating the non-equilibrium physics of a system possessing 
two adjacent transitions (cf.\ \fig\ref{fig:mixed}(middle)) might 
reveal interesting gravitational wave signatures.  
In the last case, it should be recalled that 
the transition takes place during a period in which
inflaton oscillations dominate the energy density, 
resulting in changes to the normal predictions that 
assume a radiation-dominated universe. 

%
\section*{Acknowledgements}

We are grateful to Simone Biondini, Chiara Caprini, Joachim Kopp 
and Germano Nardini for helpful discussions. 
This work was partly supported by the Swiss National Science Foundation
(SNSF) under grant 200020B-188712.

\small{
%

}

\end{document}